\documentclass[a4paper,prd,aps,preprint,tightenlines,superscriptaddress,nofootinbib,showpacs,showkeys,floatfix]{revtex4}
\usepackage{mathrsfs}
\usepackage{amsfonts}
\usepackage{array}
\usepackage{epsfig}
\usepackage{rotating}
\usepackage{multirow}

\usepackage{amsmath}    
\usepackage{verbatim}   
\usepackage{color}      
\usepackage{colordvi}
\usepackage{bm}

\usepackage{subfigure}  
\usepackage{url}
\usepackage[colorlinks = true,
            linkcolor = blue,
            urlcolor  = blue,
            citecolor = blue,
            anchorcolor = blue]{hyperref}
\usepackage{doi}

\usepackage{pstricks,rotating, graphicx}
\usepackage{cancel}

\usepackage{tablefootnote}  

\graphicspath{{./final-figs/}}

\usepackage{pslatex}
\usepackage{txfonts}
\usepackage[latin1]{inputenc}
\usepackage[T1]{fontenc}
\usepackage{appendix}

\def\muf{{\mu^{}_F}}
\def\mufs{{\mu^{\,2}_F}}
\def\mur{{\mu^{}_R}}

\def\as{{\alpha_s}}

\begin{document}
\begin{titlepage}
\thispagestyle{empty}
\noindent
PUBDB-2026-01142,\,\, P3H-26-026,\,\, TTP26-11
\hfill
April 2026 \\
\vspace{1.0cm}

\begin{center}
  {\bf \Large
Threshold Top-Quark Pair-Production: \\[0.5ex] Cross Sections and Key Uncertainties
  }
  \vspace{1.25cm}

 {\large
   M. V. Garzelli$^{\, a}$,
   G. Limatola$^{\, a}$,
   S.-O.~Moch$^{\, a}$,
   M. Steinhauser$^{\, b}$,
   O. Zenaiev$^{\, a}$
 }
 \\
 \vspace{1.25cm}
 {\it
$^a$ Universit\"at Hamburg, II. Institut f\"ur Theoretische Physik, \\
Luruper Chaussee 149, D--22761 Hamburg, Germany \\[0.5ex]
$^b$ Karlsruhe Institute of Technology (KIT),
Institut f\"ur Theoretische Teilchenphysik,
Wolfgang-Gaede Stra\ss{}e 1, D--76131 Karlsruhe, Germany\\
 }

\vspace{1.4cm}
\large {\bf Abstract}
\vspace{-0.2cm}
\end{center}

We study theoretical uncertainties in predicting top-quark pair-production near threshold at the LHC using the non-relativistic QCD framework. We include variations in the top-quark mass and width, the strong coupling $\alpha_s$, renormalization and factorization scales, and parton distribution functions, as well as uncertainties from the color-singlet and octet Green's functions that describe quasi-bound toponium formation. These uncertainties are compared with those from standard fixed-order QCD predictions, and implications for ATLAS and CMS analyses are discussed. 
For the LHC at 13 TeV center-of-mass energy, the integral of the top-quark pair invariant-mass distribution from 340 to 350~GeV is 11.67 pb with ${}^{+1.43}_{-1.47}$ pb uncertainty. The corresponding excess after subtracting the \texttt{POWHEG-BOX} result is 4.15 pb with the same uncertainties.

\end{titlepage}

\newpage
\setcounter{footnote}{0}
\setcounter{page}{1}

\section{Introduction}
\label{sec:intro}

Since the discovery of the top quark, $t{\bar t}$-pair production has become a key probe of the Standard Model and a sensitive window to new physics. 
The top-quark's large mass and extremely short lifetime prevent hadronisation, allowing perturbative quantum chromodynamics (QCD) to accurately describe its production and enabling direct access to its quantum properties 
(for recent theory and experimental reviews, see e.g. Refs.~\cite{CMS:2024irj,Schwienhorst:2022yqu}).
While most $t{\bar t}$ events at the Large Hadron Collider (LHC) are produced above threshold, the region close to and below twice the top-quark mass is increasingly accessible and phenomenologically important.

Near threshold, the $t{\bar t}$-pair behaves as a non-relativistic system. Coulomb interactions can generate quasi-bound toponium-like structures, even though the top quark decays too quickly for true bound states to form. 
These effects, described using non-relativistic QCD (NRQCD) Green's functions, enhance the cross section and modify the invariant-mass spectrum up to about 400~GeV~\cite{Fadin:1990wx,Fadin:1991zw}.
Although well studied for lepton colliders~\cite{Hoang:2000yr}, predictions for hadron colliders are less developed. 
Recent LHC measurements by the CMS~\cite{CMS:2025kzt} and ATLAS~\cite{ATLAS:2026dbe} collaborations, which reveal unexpected features in $t{\bar t}$ observables, have renewed interest in near-threshold dynamics. 

With the increasing precision of LHC data, such effects have now become measurable and have motivated a series of novel studies. These studies highlighted several key developments, including modelling of a near-threshold resonance~\cite{Fuks:2021xje}, the practicality of adding threshold effects through Green's-function reweighting of leading order simulations while preserving realistic detector acceptance and avoiding double counting~\cite{Fuks:2024yjj}, as well as the release of a new \texttt{PYTHIA} version, which now simulates $t{\bar t}$ production near threshold using leading order QCD, the Green's-function, and event generation with top-quark decays~\cite{Sjostrand:2025qez}. 
Inclusive cross sections with joint Coulomb resummation are provided in~\cite{Beneke:2012wb,Piclum:2018ndt}. 
Insights from $e^+ e^-$ threshold physics, mostly focused on the inclusive cross section~\cite{Beneke:2013jia,Beneke:2015kwa}, show that bound-state effects are strongly smeared by the top-quark width and that, with sufficiently large binning, the bound-state and continuum contributions merge into perturbatively well-behaved corrections~\cite{Nason:2025hix}.

The NRQCD perspective has been developed in~\cite{Hagiwara:2008df,Kiyo:2008bv,Sumino:2010bv}, 
and applied, for example, in searches for new physics~\cite{Maltoni:2024tul}, but further advances remain indispensable, as NRQCD is the effective theory framework for the kinematics associated with Coulomb and bound-state effects. 
Recent work in NRQCD emphasizes the importance of controlling theoretical uncertainties such as choices of scales, scheme dependence and matching~\cite{Garzelli:2024uhe}. 
The latter effects need to be taken into account because they impact Standard Model tests, top-quark mass extractions, and global fits involving parton distribution functions (PDFs) and the strong coupling $\alpha_s$. These effects will become even more relevant in the analyses of Run 3 data and at the high-luminosity LHC (HL-LHC).

Given the CMS~\cite{CMS:2025kzt} and ATLAS~\cite{ATLAS:2026dbe} measurements, we aim to establish a well-defined procedure to compute the excess cross section near threshold due to $t{\bar t}$ bound-state effects, which requires a shared reference framework to compare theoretical approaches and experimental analyses. 
Key priorities include establishing common binning and fiducial phase-space definitions, as well as improving the understanding of modelling systematics in the low $t{\bar t}$ invariant-mass region.
In this context, the purpose of the present work is to revisit predictions for the $t{\bar t}$ invariant-mass distribution near threshold and to provide a comprehensive assessment of theoretical uncertainties, including variations in the top-quark mass and width, PDFs and $\alpha_s$, renormalization and factorization scales, and uncertainties in the QCD potential entering the Green's-function calculation.

After outlining the theoretical framework in Sec.~\ref{sec:theo}, we present predictions with uncertainties in Sec.~\ref{sec:pred}, discuss implications for current experimental analyses in Sec.~\ref{sec:impli}, and conclude in Sec.~\ref{sec:conclu}. 
To support ongoing analyses, 
we provide tables with predictions for different input parameters.

\section{Theory framework}
\label{sec:theo}
In this section, we review the theoretical ingredients relevant for describing $t{\bar t}$ production near threshold. 
These include the NRQCD framework used to capture the non-relativistic dynamics of the quasi-bound $t{\bar t}$ system, as well as fixed-order perturbative calculations and their matching to a parton-shower framework.
In the experimental analyses, the latter provides a QCD description of $t{\bar t}$ production at some fixed order in perturbation theory matched to a parton shower, possibly incorporating spin-correlated top-quark decays, radiation in production and decay, and a realistic modelling of hadronization and underlying-event effects.

Let us summarize the theoretical framework, following Ref.~\cite{Kiyo:2008bv}.
We study the production of a $t\bar{t}$ pair in a quasi bound-state with invariant-mass $M_{t\bar{t}}$, denoted as  
$T\equiv {}^{2S+1}L_{J}^{[1,8]}$, where $S,~L,~J$ are the spin, orbital, and total angular momentum of the state, respectively and the superscripts $[1,8]$ indicate singlet and octet colour configurations.
The production rate of $T$ in hadron collisions with momenta $P_1$ and $P_2$ factorises at the scale $\mu_F$ as a convolution 
\begin{equation}
\label{eqn:QCDmasterformula}
 M_{t\bar{t}} {{\rm d}\sigma_{P_1 P_2\to T}\over {\rm d}M_{t\bar{t}}}(S, M_{t\bar{t}}^2) 
 \,=\,
 \sum_{i,j} \int_\rho^1 {\rm d}\tau\,
  \bigg[\frac{{\rm d}{\cal L}_{ij}}{{\rm d}\tau}\bigg](\tau,\mufs)\, 
  M_{t\bar{t}}\frac{{\rm d}\hat\sigma_{ij\to T}}{{\rm d}M_{t\bar{t}}}(\hat s,M_{t\bar{t}}^2,\mufs)\,,
\end{equation}
of the parton level production rate ${\rm  d}\hat\sigma_{ij\to T}/{{\rm d}M_{t\bar{t}}}$ and the parton luminosity

\begin{equation}
\label{eqn:lum}
\biggl[\frac{\rm d\mathcal{L}_{ij}}{\rm d\tau}\biggr](\tau,\mufs) \,=\,
    \int_0^1 {\rm d} x_1 {\rm d} x_2\, f_i(x_1,\mufs) f_j(x_2,\mufs)\,
    \delta(\tau-x_1 x_2)\, ,
\end{equation}
where $i,j$ denote the partons inside the hadrons, with the distribution functions $f_i, f_j$.
Here $S$ and $\hat{s}$ are the hadronic and partonic center-of-mass energies, respectively. 
Useful dimensionless variables are $\tau=\hat{s}/S$ and $\rho=M_{t\bar{t}}^2/S$.

\subsection{Fixed-order QCD calculations matched to parton shower}
\label{sec:fixed}

The state-of-the-art of theoretical predictions for the $M_{t\bar{t}}$ differential distribution in Eq.~\eqref{eqn:QCDmasterformula}
considering stable ${t\bar{t}}$ production is represented by next-to-next-to-leading order (NNLO)
QCD calculations~\cite{Czakon:2015owf,Catani:2019hip}, combined with next-to-leading order (NLO) electroweak
corrections~\cite{Czakon:2017wor} and additional resummation of threshold logarithms~\cite{Czakon:2019txp}.
Furthermore, fixed-order predictions have been matched with parton showers within various frameworks. 
Event generation at NNLO, combined with a parton shower has been implemented in a consistent event generator using the \texttt{MiNNLO$_{\texttt{PS}}$}~\cite{Monni:2019whf,Mazzitelli:2020jio} framework.
Following the  set-up of the CMS analyses, predictions at NLO in QCD, obtained from the \texttt{hvq}~\cite{Frixione:2007nw} generator within the \texttt{POWHEG-BOX}~\cite{Nason:2004rx, Frixione:2007vw, Alioli:2010xd} have been employed to simulate ${t\bar{t}}$ production and decay.\footnote{A chain of simulations more or less equivalent to  \texttt{POWHEGBOX-hvq} can be built upon NLO QCD predictions for $t{\bar t}$ production provided in 
\texttt{MC@NLO}~\cite{Frixione:2007zp} and \texttt{MadGraph5\_aMC@NLO}~\cite{Artoisenet:2012st,Alwall:2014hca}.} 
This approach utilizes the narrow width approximation (NWA), which simplifies the modeling by treating the top-quark as an on-shell particle with a decay width $\Gamma_t$ that is significantly smaller than its mass $m_t$. 
Under this approximation, the production and decay processes are factorized, and the full cross section is computed as the product of the on-shell production cross section in Eq.~\eqref{eqn:QCDmasterformula} and the top-quark decay rate.

The finite width of the top-quark induces a smearing effect around the production threshold at approximately twice the top mass $2m_t$,  thereby softening the sharp onset expected for stable top-quarks. This effect is further amplified by the treatment of top-quark decay within \texttt{POWHEG-BOX}, which permits decay products to be produced with invariant masses below the nominal threshold, leading to predictions that fill bins below $2m_t$.
For example, the CMS collaboration employs this methodology in top-quark event modeling. 
The NWA neglects off-shell contributions and quantum interference effects between production and decay, which have long been studied at NLO in QCD~\cite{Denner:2010jp,Bevilacqua:2010qb,Denner:2012yc}. 
More recently, their matching to parton shower~\cite{Jezo:2016ujg,Jezo:2023rht} as well as progress towards NNLO~\cite{Buonocore:2025fqs} has been achieved.

Near the production threshold, electroweak NLO and non-factorizable QCD corrections are negligible, and off-shell effects are also minimal due to the constrained phase space. We employ the \texttt{hvq} generator and \texttt{POWHEG-BOX} predictions in accordance with CMS analysis procedures. For fixed-order calculations assuming stable top quarks, we utilize the \texttt{MATRIX} framework~\cite{Catani:2019hip} as discussed in \cite{Garzelli:2023rvx}.

\subsection{NRQCD predictions}
\label{sec:nrqcd}

The top-quark velocity in the center-of-mass frame is given by  $\beta_t=\sqrt{1-4m^2_t/M^2_{t\bar{t}}}$.  
In the threshold region, where $\beta_t \ll1$, NRQCD techniques allow one to factorize the short-distance production of a state $T$ from the long-distance quasi bound-state dynamics of the produced $t{\bar t}$ pair, leading to
\begin{equation}
    \label{eqn:NRQCDformula}
    M_{t\bar{t}}\frac{{\rm d}\hat{\sigma}_{ij\to T}}{{\rm d}M_{t\bar{t}}} \,=\,
    F_{ij\to T}(\hat{s},M^2_{t\bar{t}},\mufs)\, 
    \frac{1}{m_t^2}\, 
    {\rm Im}~G^{[1,8]}(M_{t\bar{t}}+i\Gamma_t)\, ,
\end{equation}
where $m_t$ and $\Gamma_t$ are the top-quark's mass and width, respectively.
$F_{ij\to T}$ is the short-distance hard function and $G^{[1,8]}$ are the zero-distance Coulomb Green's functions, which are solutions of Schr{\"o}dinger equations with Coulomb-like QCD potentials $V^{[1,8]}$. 
These potentials describe an attractive interaction in the colour-singlet $(V^{[1]})$  configuration and a repulsive one in the colour-octet $(V^{[8]})$ configuration.
Since the production of a state $T$ takes place at the scale $\mu\sim m_t$, the hard functions $F_{ij\to T}$ are evaluated up to next-to-leading order (NLO) in $\alpha_s$~\cite{Kuhn:1992qw,Petrelli:1997ge}
\begin{align}
\label{eqn:freeTrate}
F_{ij\to T}(\hat{s},M^2_{t\bar{t}},\mufs)=\mathcal{N}_{ij\to T}\frac{\pi^2\alpha_s^2(\mur)}{3 \hat{s}}\biggl[\biggl(1+\frac{\alpha_s(\mur)}{\pi}\mathcal{C}_h\biggr)~\delta_{ij\to T}\delta(1-z)+\frac{\alpha_s(\mur)}{\pi}\biggl(\mathcal{A}_c(z)+\mathcal{A}_{nc}(z)\biggr)\biggr]\, ,
\end{align}
where the $\mathcal{C}_h$ terms capture the contributions arising from the hard corrections, and $\mathcal{A}_{c~(nc)}$ contain the real corrections due to the (non-) collinear QCD emissions. 

Furthermore, we have also incorporated the resummation of threshold-enhanced logarithms associated with soft gluon emissions up to next-to-leading logarithmic (NLL) accuracy, along the same lines adopted in~\cite{Kiyo:2008bv,Garzelli:2024uhe}.
This resummation, applied to the three most relevant contributions to cross sections, corresponding to the channels $gg\rightarrow {}^1S_0^{[1]},\,gg\rightarrow {}^1S_0^{[8]} ,\,q\bar{q}\rightarrow {}^3S_1^{[8]}$ (which already contribute at $\mathcal{O}(\as^2)$), results in an enhancement of the predictions by a few percent.  
The methodology follows the approach discussed in Section~4 of Ref.~\cite{Kiyo:2008bv} and is also employed in Ref.~\cite{Garzelli:2024uhe}.

Since the non-relativistic momentum scale is $m_t\beta_t\sim m_t\alpha_s$ and $\Gamma_t\gg \Lambda_{\rm QCD}$, the Green functions can be computed perturbatively at NLO in NRQCD. At this order, the Green function and the convolution $\mathcal{L}\otimes F$ are separately independent of the renormalisation scale $\mu_R$, and can therefore be analysed independently.


\section{Predictions}
\label{sec:pred}

\subsection{Setup and estimate of uncertainties}

For our central predictions, we focus on proton-proton collisions at the LHC with a center-of-mass energy of $\sqrt{S} = 13\,\mathrm{TeV}$. 
We use the top-quark on-shell (pole) mass $m_t = 172.5~\mathrm{GeV}$ as the central input scale, with the renormalization and factorization scales both set equal to this value 
($\mu_{R,0}=\mu_{F,0}= m_t$). 
The PDFs are taken from the NNPDF3.1\_nnlo\_as\_118 set~\cite{NNPDF:2017mvq}, with the associated value of the strong coupling $\alpha_s(M_Z)=0.118$.
The top-quark decay width is set to $\Gamma_t = 1.36\,\mathrm{GeV}$~\cite{CMS:2014mxl}.

In the NRQCD framework, the Green's functions depend on the top-quark mass defined in different schemes. 
We use a potential scheme mass $m_t^{PS} = 170.1\,\mathrm{GeV}$, 
which is compatible with the quoted value for on-shell mass $m_t^{OS}$.
The potential scheme mass is designed to improve the convergence of the perturbative series by subtracting certain infrared-sensitive effects present in the pole mass scheme. 
In summary, the parameters are chosen to reflect the conventional pole mass scheme for the top quark, while acknowledging the relation to the potential scheme, and employing a practical approach in the Green's function calculations.

As we will demonstrate in the following, the parametric uncertainties listed below ($\alpha_s$, $m_t$, $\Gamma_t$ and PDFs) are evaluated based on NLO QCD predictions, with the expectation that the size of the uncertainty bands remains roughly the same when including NLL resummed corrections. 
For the central scale choice $\mu_{R,0}=\mu_{F,0}=m_t$ adopted in this analysis, the NLO and NLO+NLL predictions differ only by a few percent in the threshold region. In general, the inclusion of resummation substantially reduces the sensitivity to the hard scale variations~\cite{Kiyo:2008bv,Garzelli:2024uhe}.
We perform a comprehensive study of uncertainties by varying key parameters within their typical ranges. 

\noindent
{\bf{Renormalization and factorization scale variation}}:
Regarding the hard renormalization and factorization scales, $\mu_R$ and $\mu_F$, we vary both simultaneously, usually up and down by a factor of two relative to the central scale $\mu_{R,0}=\mu_{F,0}=m_t$, where $m_t = 172.5\,\mathrm{GeV}$.

\noindent
{\bf{Soft-scale variation and Green's function at higher-orders:}}
We vary the soft scale $\mu_s$, chosen around $30$\,GeV, within the range 20 to 40\,GeV; this scale is used in the computation of the Green's functions, with the central value matching the approach used in \cite{Kiyo:2008bv}. 
Furthermore, we estimate the effect due to higher order
corrections to the Green's functions by using the Coulomb
approximation at NNLO.\footnote{Note that the full NNLO Green's function is
divergent; this aspect is yet to be fully implemented.}
As final estimate of the uncertainty we use the envelope
of the two approaches.

\noindent
{\bf{Strong coupling:}}
In the case of the strong coupling $\alpha_s$, we compare results obtained with the NNPDF3.1\_nnlo\_as\_0118 set (with $\alpha_s(M_Z) = 0.118$) to those from variants with $\alpha_s(M_Z) = 0.116$ and $0.120$, specifically the sets NNPDF3.1\_nnlo\_as\_0116 and NNPDF3.1\_nnlo\_as\_0120, keeping  $\alpha_s$ correlated with the PDFs.

\noindent
{\bf{Top-quark mass:}}
Regarding the top-quark on-shell mass $m_t$, we analyze the effect of $\pm 1\,\mathrm{GeV}$ variations around the central value, comparing results for $m_t=172.5\,\mathrm{GeV}$ with those for $m_t=171.5\,\mathrm{GeV}$ and $173.5\,\mathrm{GeV}$. 

\noindent
{\bf{Top-quark width:}}
For the top width $\Gamma_t$, we vary it by $\pm 0.05\,\mathrm{GeV}$ around its central value of $1.36\,\mathrm{GeV}$, and the impact of this variation is confirmed to be small.

\noindent
{\bf{PDF variation:}}
We also consider other PDF sets, such as ABMP16als118\_5\_nnlo~\cite{Alekhin:2017kpj}, CT18\_NNLO~\cite{Hou:2019efy}, MSHT20\_nnlo\_as\_118~\cite{Bailey:2020ooq}, NNPDF3.0\_nnlo\_as\_118~\cite{NNPDF:2014otw}, and NNPDF4.0\_nnlo\_as\_118~\cite{NNPDF:2021njg}, each with their respective $\alpha_s(M_Z)$ and $m_t$ values, see 
Sec.~\ref{sec:PDFunc} for details. 
We estimate PDF uncertainties using the envelope of different sets with the same $\alpha_s(M_Z)$ and $m_t$ values, and analyze the uncertainty band from the default NNPDF3.1\_nnlo\_as\_118 set. 
Finally, we evaluate combined uncertainties by performing simultaneous variations of $\alpha_s(M_Z)$, $m_t$, and PDFs, considering correlations among these parameters. For example, the comparison of the default ABMPtt\_5\_nnlo set~\cite{Alekhin:2024bhs} at $\alpha_s(M_Z)=0.115$ and $m_t = 170.2\,\mathrm{GeV}$ to specialized sets with different values provides insight into the combined effect on our predictions.


\subsection{Scale variation}

\begin{figure}[t]
  \begin{center}
  \includegraphics[width=0.49\textwidth]{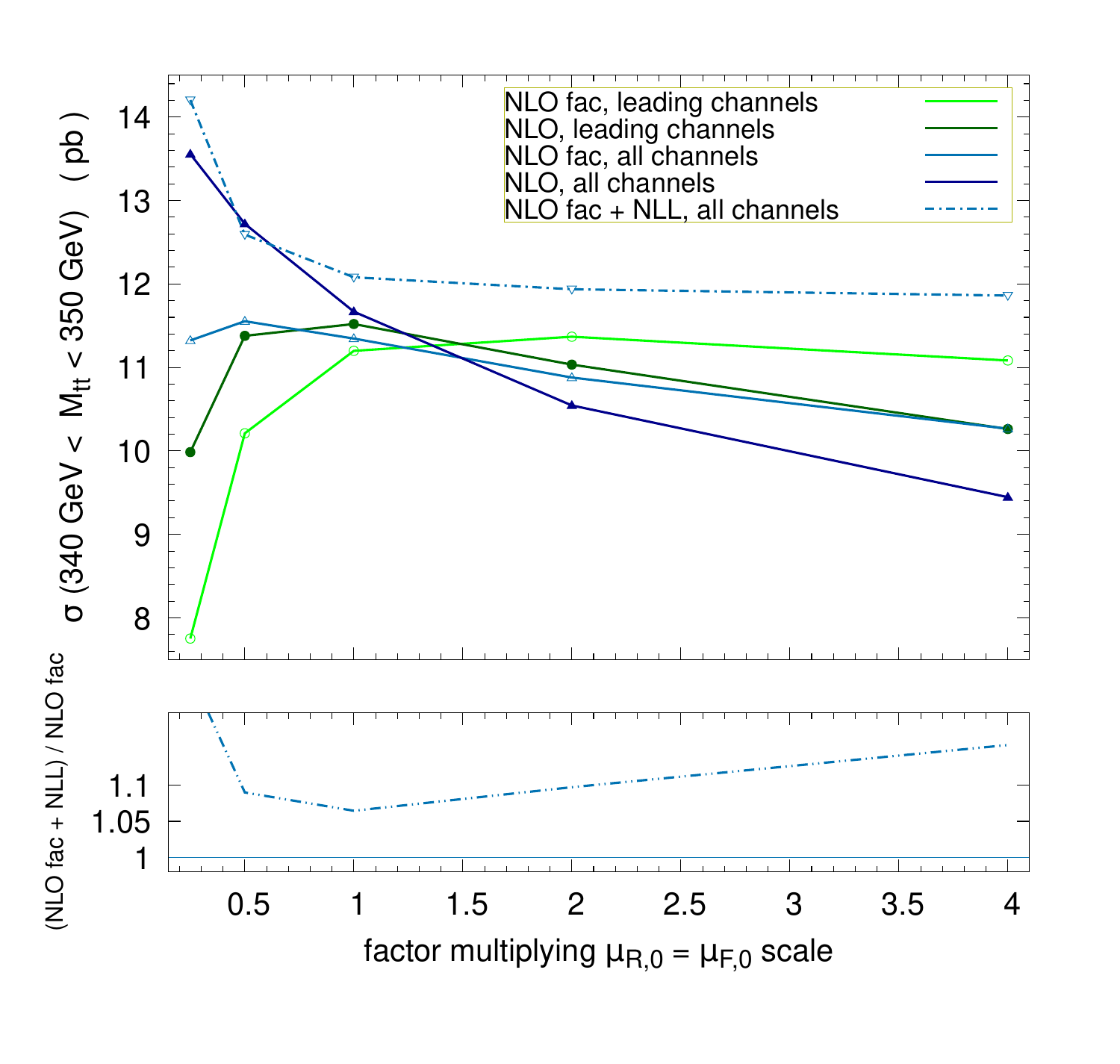}
  \includegraphics[width=0.49\textwidth]{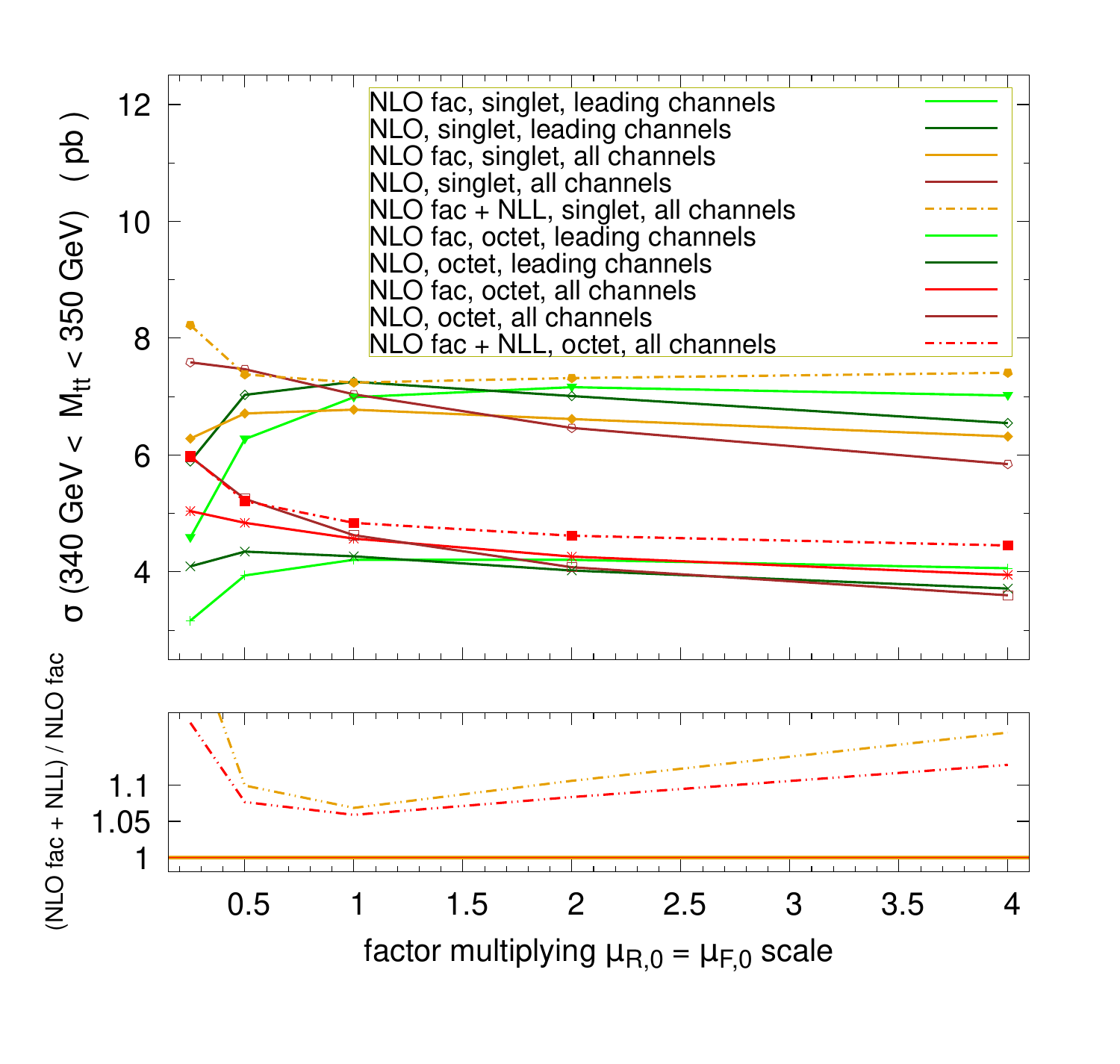}
  \caption{\label{fig:1hardscale} 
Cross section in the $M_{t\bar{t}} \in [340, 350]\, \mathrm{GeV}$ bin as a function of hard scale variations taken as multiples of $\mu_{R,0}=\mu_{F,0}=172.5$ GeV. The curves are obtained through the implementation of $F_{ij\to T}$ in Eq.~\eqref{eqn:freeTrate} and its factorised version ("fac"), including the hard corrections~(see footnote~\ref{note:fac}). In the left panel the singlet and octet contributions are summed together, while in the right panel the singlet and octet contributions are plotted separately. We also show the contribution only from the leading production channels. In both the panels, the dashed curves refer to the NRQCD cross section with NLL resummation included for the factorized $F_{ij\to T}$.}
 \end{center}
\end{figure}
In the left panel of Fig.~\ref{fig:1hardscale}, we show the cross section in Eq.~\eqref{eqn:NRQCDformula} integrated over $M_{t\bar{t}}$ within the $[340,350]$ GeV range, as a function of hard scale variations, taken as multiples of the central scale $\mu_{R,0}=\mu_{F,0}=m_t$. 
The contributions from all the $t\bar{t}$ production subprocesses and for the three leading ones are separately displayed.
The results are shown for two different definitions of the functions $F_{ij\to T}$, namely the non-factorized expression in Eq.~\eqref{eqn:freeTrate} and the factorized expression\footnote{\label{note:fac}
The factorized version pulls out the hard corrections $\biggl(1 + \frac{\alpha_s(\mu_R)}{\pi} \mathcal{C}_h \biggr)$ in Eq.~\eqref{eqn:freeTrate} as an overall multiplicative factor, cf. Eq.~(6) in Ref.~\cite{Kiyo:2008bv}.}
(also used in our previous work~\cite{Garzelli:2024uhe}). 
At smaller scales, such as $m_t/2$ and $m_t/4$, neglecting the subleading channels reduces the cross section by about 10\% and nearly 30\%, respectively. This highlights the importance of considering all channels for accurate predictions.
Additionally, at lower scales, the predictions based on the non-factorized formula~\eqref{eqn:freeTrate} tend to be larger than those from the factorized approach. 
We have identified the hard scale $\mu_{R,0}=\mu_{F,0}= m_t$ as providing the smallest variation among the different predictions, which is why we chose it as the central scale for subsequent analyses. 
Furthermore, this scale choice minimises the impact of NLL resummation over the NLO NRQCD, as demonstrated in the bottom panels of Fig.~\ref{fig:1hardscale}, where we show the ratios of the central  NLO+NLL resummation prediction compared to the NLO NRQCD result, using the factorized form of $F_{ij\to T}$.  
Similar trends are observed when analyzing and plotting the singlet and octet components separately, as shown in the right panel of Fig.~\ref{fig:1hardscale}.
It is important to note that the estimates of other parametric uncertainties ($\alpha_s$, $m_t$, $\Gamma_t$ and PDFs) 
shown below are based solely on NLO variations, since we do not expect the inclusion of NLL corrections to significantly change the size of these uncertainty bands.

%
\begin{figure}[t]
  \begin{center}
  \includegraphics[width=0.49\textwidth]{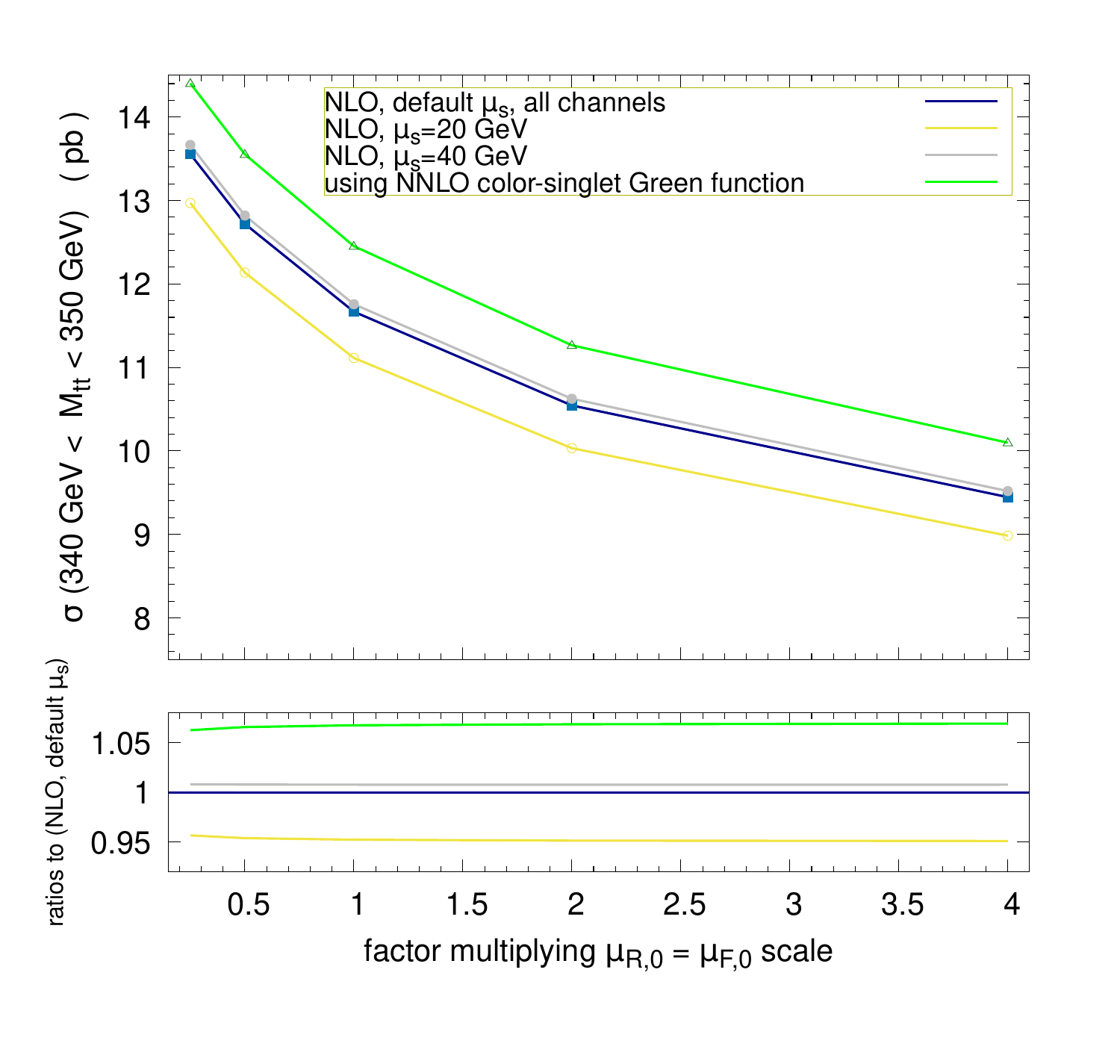}
  \includegraphics[width=0.49\textwidth]{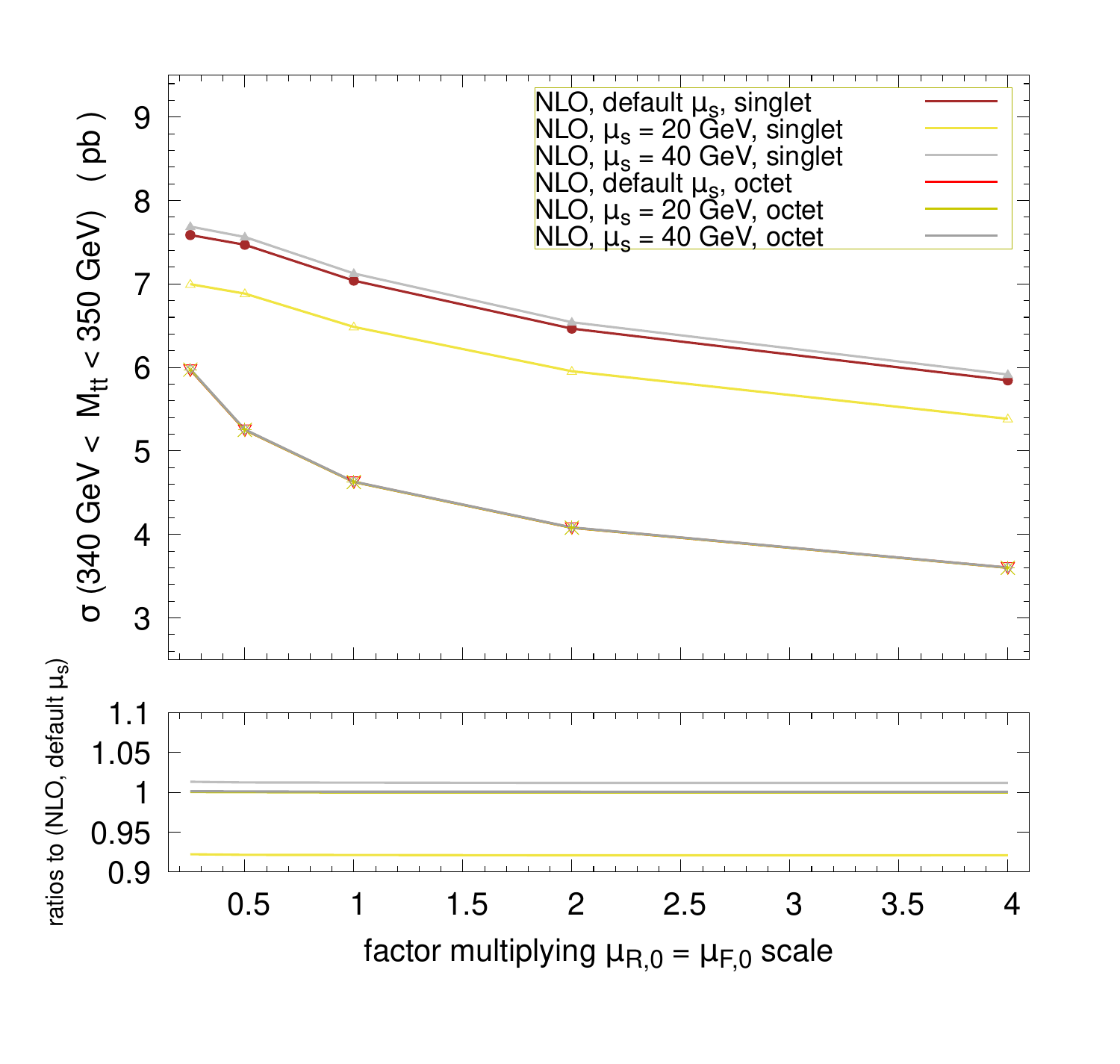}
  \caption{\label{fig:1softscale} 
Cross section in the $M_{t\bar{t}} \in [340, 350]\, \mathrm{GeV}$ bin with $\mu_s = 20\,,32\,, 40\, \mathrm{GeV}$ 
as a function of hard scale variations taken as multiples of $\mu_{R,0}=\mu_{F,0}=172.5$ GeV. 
In the left panel the contributions for the colour singlet and octet final states are combined together, while in the right panel they are displayed separately. The light-green curve in the left panel has been obtained using the Green function up to NNLO.
In the right panel, the color-octet curves are hardly distinguishable because they almost lie on top of each other.
}
 \end{center}
\end{figure}
In Fig.~\ref{fig:1softscale} (left panel), we present the dependence of the cross section in the region  $340 < M_{t\bar{t}} < 350\, \mathrm{GeV}$ on the soft scale $\mu_s$, which exclusively enters the Green's functions. 
Since at NLO in NRQCD $F_{ij\to T}$ and the Green's functions are separately renormalization-group invariant, $\mu_s$ may be chosen independently of the hard scale $\mu_{R,0}=\mu_{F,0}$ appearing in $F_{ij\to T}$ in Eq.~\eqref{eqn:freeTrate}.
The multiple Coulomb-like virtual gluon exchanges encoded in the Green's functions probe progressively lower momentum scales and therefore correspond to larger values of $\alpha_s$ than those associated with the hard-scattering vertex at scale $\mu_{R,0}=\mu_{F,0}=m_t$. 
In line with previous studies, we adopt the prescription of Ref.~\cite{Beneke:2005hg} for the central value of $\mu_s$, yielding approximately $\mu_s \approx 32\, \mathrm{GeV}$. 
Varying  $\mu_s$ in the range $\mu_s \in [20,\, 40]\, \mathrm{GeV}$ induces a change of roughly $+3\%$ and $-9\%$ in the peak region, with the smallest cross section obtained for the smallest scale. 
As shown in the bottom panels of Fig.~\ref{fig:1softscale}, this behaviour remains essentially unchanged when varying the hard scale $\mu_{R,0}=\mu_{F,0}=m_t$ by overall multiplicative factors as demonstrated in Fig.~\ref{fig:1hardscale}. 
The sensitivity to $\mu_s$ arises almost entirely from the color-singlet contribution; the color-octet contribution is nearly insensitive to the soft scale settings, as shown in Fig.~\ref{fig:1softscale} (right panel), where the effects of $\mu_s$ variation are displayed separately for the singlet and octet components.
In the left panel of Fig.~\ref{fig:1softscale} we also show the
predictions where we use the NNLO result for the Coulomb Green's function. 
We observe a positive shift of about 10\% independent from the choices of $\mu_{R,0}$ and $\mu_{F,0}$.


\subsection{Strong coupling variation}

\begin{figure}[t]
  \begin{center}
  \includegraphics[width=0.49\textwidth]{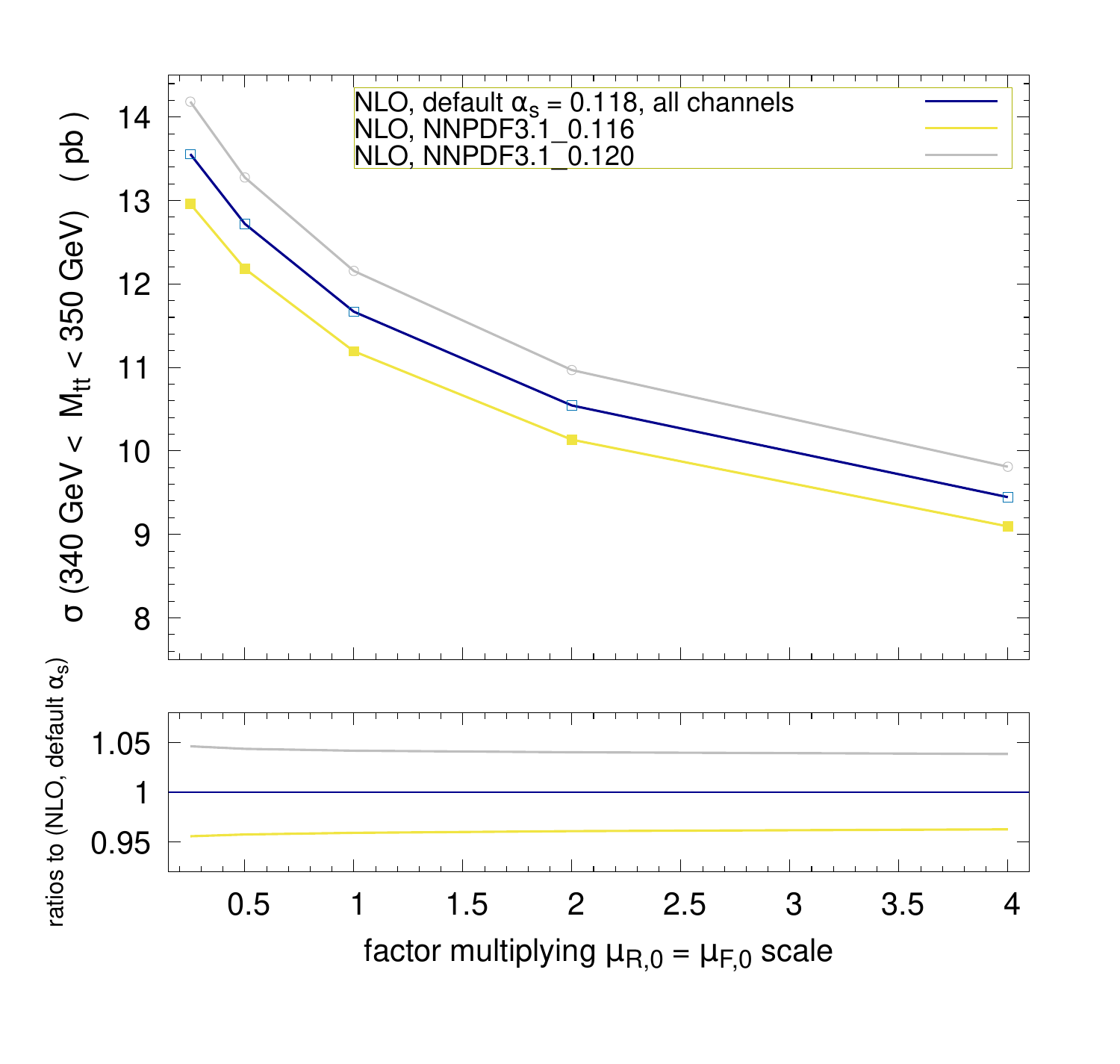}
  \caption{\label{fig:1as} 
Cross section in the $M_{t\bar{t}} \in [340, 350]\, \mathrm{GeV}$ bin using Eq.~\eqref{eqn:freeTrate}
as a function of hard scale variations taken as multiples of $\mu_{R,0}=\mu_{F,0}=172.5$ GeV, varying $\alpha_s(M_Z) =  0.118\pm0.002$ through PDF choices (see text), and keeping all the other parameters fixed.
  }
 \end{center}
\end{figure}
\begin{figure}[t]
  \includegraphics[width=0.49\textwidth]{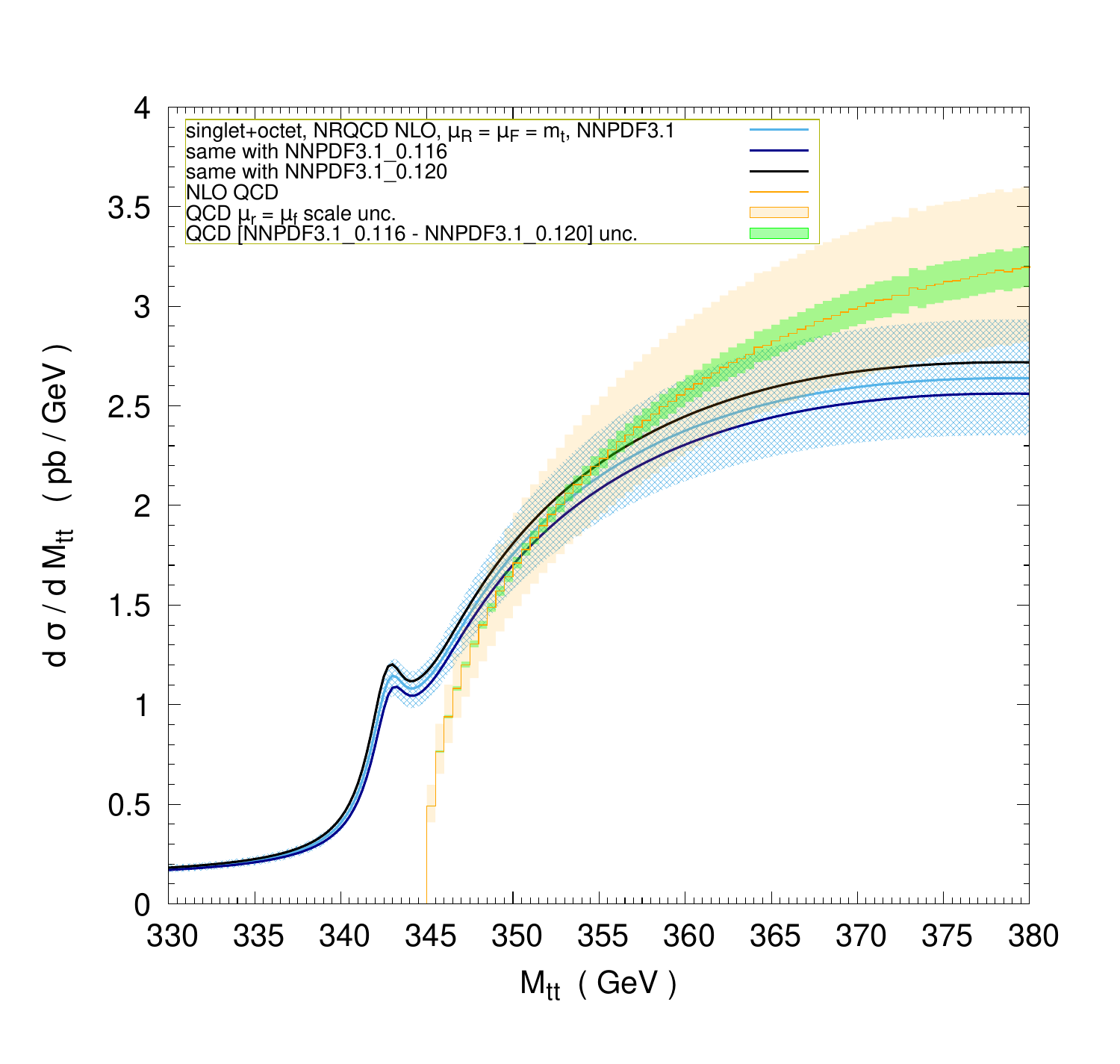}
  \includegraphics[width=0.49\textwidth]{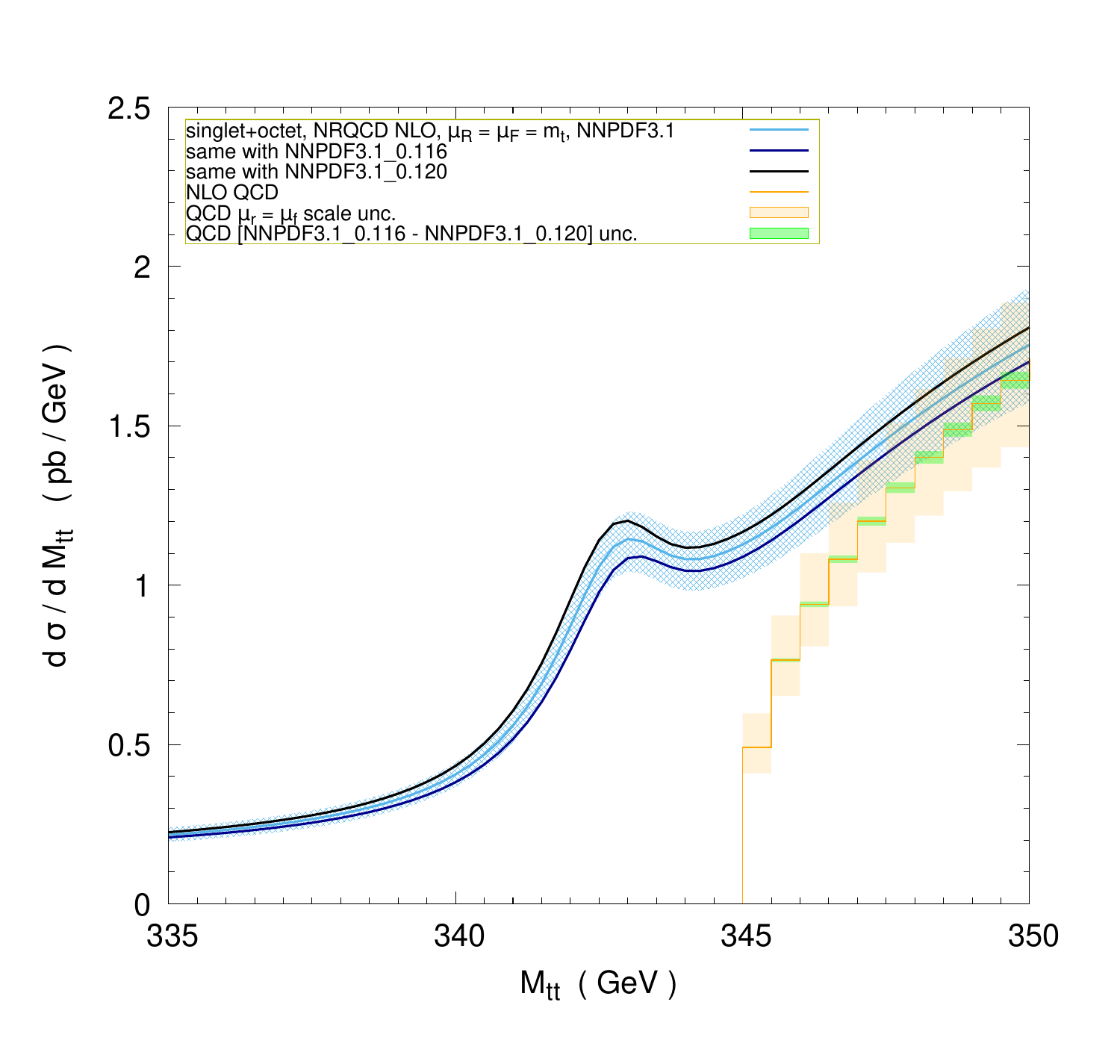}
  \caption{\label{fig:2as} 
Left panel: NRQCD predictions for $d\sigma/dM_{t\bar{t}}$ obtained with the central scale choice $\mu_{R,0} = \mu_{F,0} = m_t$ and the NNPDF3.1 PDF set using its default value $\alpha_s(M_Z) = 0.118$, compared with the ones using $\alpha_s(M_Z) = 0.116$ and $0.120$. The hard scale variation uncertainty (light blue) refers to our default setup. 
In orange the same comparison has been displayed for fixed-order NLO predictions for $t\bar{t}$ pair production.
Right panel: zoom of the same comparison in the threshold region.
    }
\end{figure}
Besides studying scale-dependence effects, it is also important to assess the intrinsic uncertainty associated with the strong coupling $\alpha_s$ itself. This uncertainty is typically parametrized through variations of $\alpha_s(M_Z)$. 
Given the different extraction methods and the spread of current determinations summarized by the Particle Data Group~\cite{ParticleDataGroup:2024cfk}, we adopt the conservative choice $\alpha_s(M_Z) = 0.118 \pm 0.002$ for most of our predictions and quote the corresponding variations as the $\alpha_s(M_Z)$ uncertainty. 

As shown in Fig.~\ref{fig:1as}, this uncertainty amounts to a few percent and remains approximately constant under variations of the hard scale $\mu_{R,0}=\mu_{F,0}$ by multiples of $m_t$.
Our central predictions are based on the NNPDF3.1 PDF set, which uses $\alpha_s(M_Z)$ as an external input in the PDF fits. 
The NNPDF collaboration also provides PDF sets corresponding to $\alpha_s(M_Z) = 0.116$ and $0.120$, which we use in Fig.~\ref{fig:1as}, thereby accounting for correlations between $\alpha_s(M_Z)$ and the PDFs. 
In the bottom panel of Fig.~\ref{fig:1as}, we show the ratio of these predictions to the one with the default choice of $\alpha_s(M_Z)$. 
The other PDF sets considered in this work likewise assume the same central $\alpha_s(M_Z)$ value, except for the ABMP sets, estimated through a simultaneous fit of the PDFs, the values of $\alpha_s(M_Z)$, and heavy-quark masses. 
The resulting values are around $\alpha_s(M_Z) = 0.1150$, which is not included in the variation band used in Fig.~\ref{fig:1as}. 

Figure~\ref{fig:2as} illustrates the impact of varying $\alpha_s(M_Z)$ on the predicted $M_{t\bar{t}}$ differential cross sections. 
The differences between the three $\alpha_s(M_Z)$ values remain moderate across the full $M_{t\bar{t}}$ spectrum and are generally smaller than, or comparable to, the scale-variation uncertainties of the default setup. 
In the threshold region, highlighted in the right panel in Fig.~\ref{fig:2as}, the dependence on $\alpha_s(M_Z)$ becomes slightly more pronounced but remains within the overall theoretical uncertainty. 
This demonstrates that the sensitivity of the predictions to the current $\alpha_s(M_Z)$ uncertainty is limited and does not significantly alter the qualitative behaviour of the cross section near threshold.

\subsection{Top-quark mass variation}

\begin{figure}[t]
  \begin{center}  \includegraphics[width=0.49\textwidth]{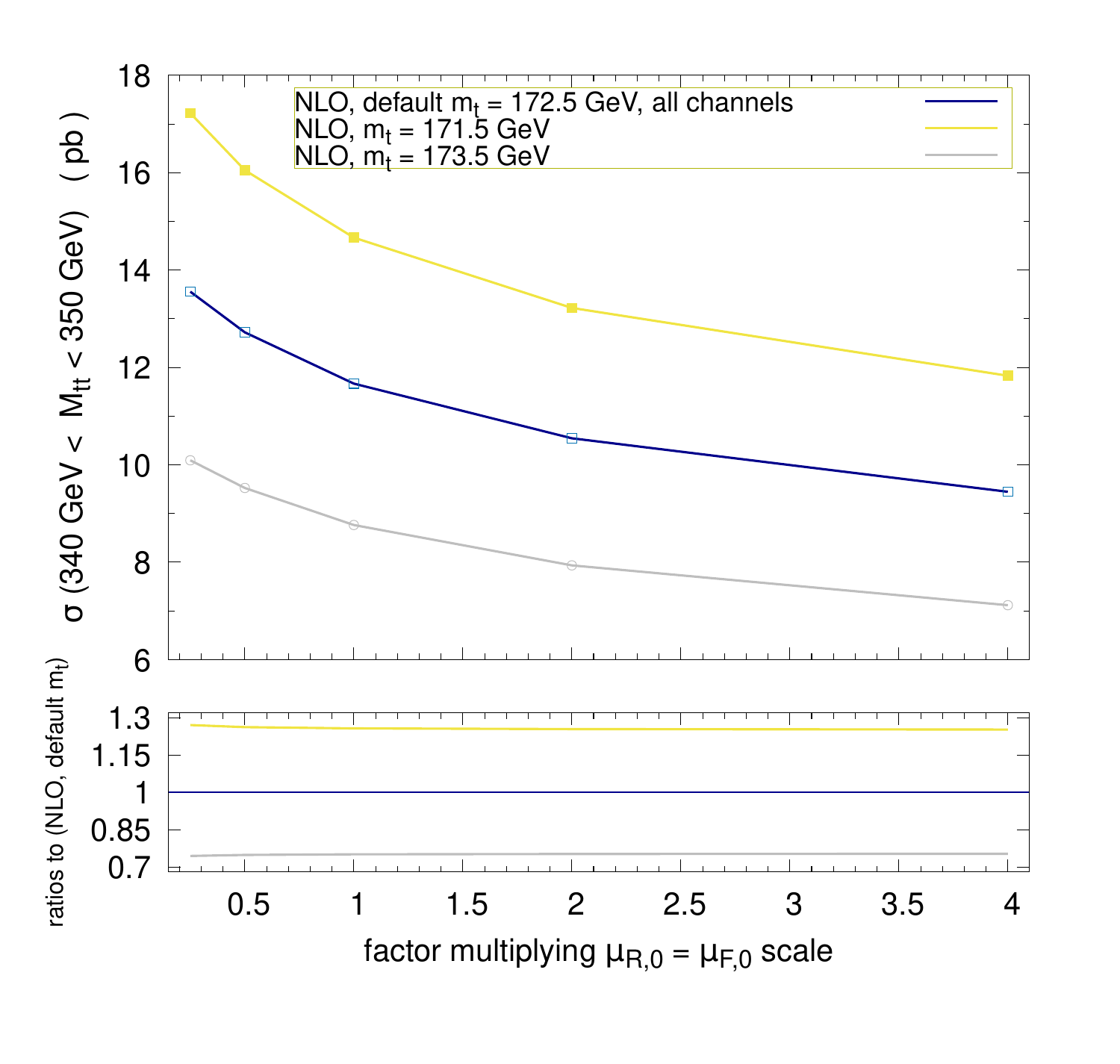}
  \caption{\label{fig:1mt} 
Cross section in the $M_{t\bar{t}} \in [340, 350]\, \mathrm{GeV}$ bin using Eq.~\eqref{eqn:freeTrate}
as a function of hard scale variations taken as multiples of $\mu_{R,0}=\mu_{F,0}=172.5$ GeV, 
varying $m_t=172.5\pm 1$ GeV and keeping all the other parameters fixed.
 }
 \end{center}
\end{figure}
\begin{figure}[h!]
  \includegraphics[width=0.49\textwidth]{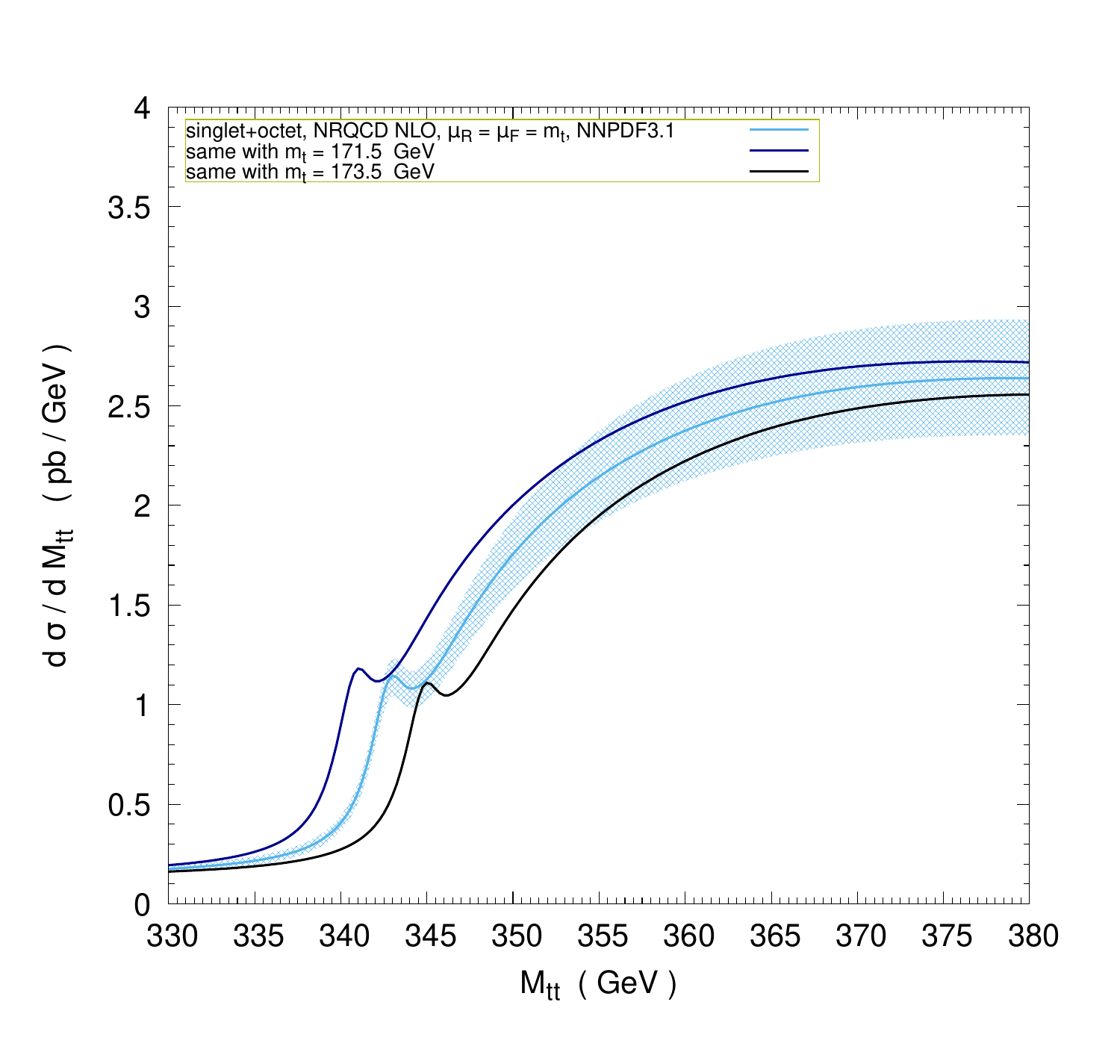}
  \includegraphics[width=0.49\textwidth]{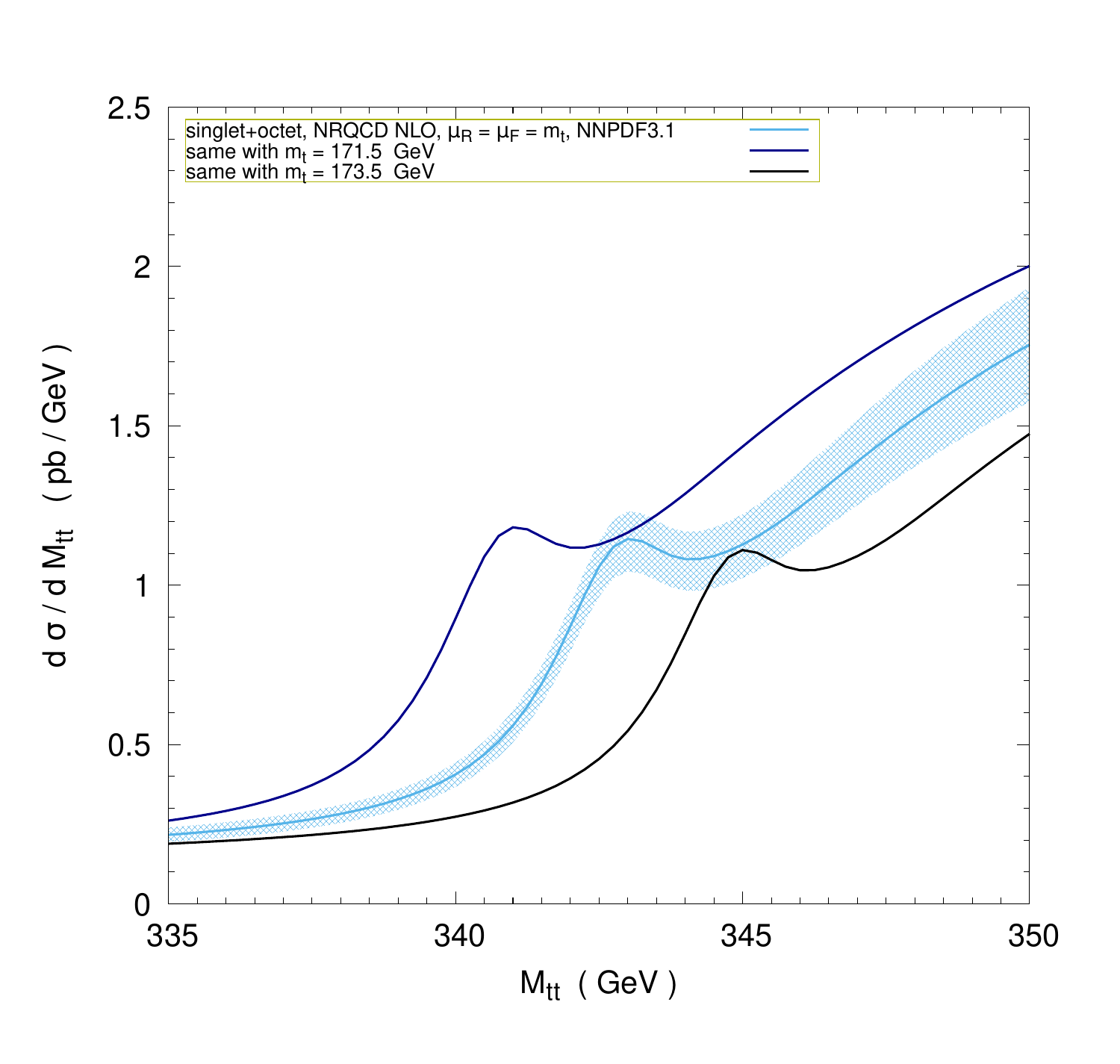}
  \caption{\label{fig:2mt} 
Left panel: predictions for $d\sigma/dM_{t\bar{t}}$ with  $\mu_R = \mu_F = m_t$ using the NNPDF3.1 PDF set with its default $\alpha_s(M_Z)$ and $m_t = 172.5\, \mathrm{GeV}$, compared with results for $m_t = 171.5\, \mathrm{GeV}$ and $173.5\, \mathrm{GeV}$.  
The light-blue band shows the hard scale uncertainty of the default setup. Right panel: zoom of the same comparison in the threshold region.
    }
\end{figure}
In Fig.~\ref{fig:1mt} we study the effect of varying the top-quark mass by $\pm 1\,\mathrm{GeV}$ around the central pole mass value $m_t = 172.5\,\mathrm{GeV}$, which is the value used by the CMS collaboration in their toponium analyses~\cite{CMS:2025kzt}. 
In this comparison we neglect 
correlations between $m_t$ and other input parameters, in particular the 
PDFs and $\alpha_s(M_Z)$. 
The bottom panel of Fig.~\ref{fig:1mt} displays the ratio of these predictions compared to the one with the default choice of $m_t$.
The impact on the cross section in the region $340 < M_{t\bar{t}} < 350\, \mathrm{GeV}$ is sizable, around $\pm 25\%$, with 
larger values of $m_t$ leading to smaller cross sections. 
This behaviour arises because the integration limits ($340$ and $350\,\mathrm{GeV}$) are kept fixed, while the position of the peak in the color-singlet contribution shifts with the top-quark mass. 
The peak typically occurs near $M_{t\bar{t}} \simeq 2 m_t - B$, where $B$ is the binding energy of the toponium quasi-bound 
state, approximately 2\,GeV.

Figure~\ref{fig:2mt} illustrates the dependence of the predicted 
$M_{t\bar{t}}$ differential distribution on the value of the top-quark mass. 
As shown in the left panel, shifting $m_t \pm 1\,\mathrm{GeV}$ around the central value of 172.5\,GeV leads to noticeable changes in the distribution, with the larger mass yielding a visibly reduced prediction. 
These variations are significantly larger than the hard scale uncertainty band of the default setup, shown in light blue. 
The right panel provides a zoom into the threshold region, where the sensitivity to $m_t$ becomes particularly pronounced due to the shift in the position of the color-singlet peak. 
This behaviour highlights the strong impact of the top-quark mass on the shape and normalization of the $M_{t\bar{t}}$ spectrum near threshold.

Among the parametric uncertainties considered, the variation of $m_t$
produces by far the largest effect near the peak. 
With improved  experimental precision in the future, such sensitivity could enable determinations of the top-quark mass from differential $t\bar{t}$ distributions, provided bin sizes of order 5 to 10\,GeV become achievable. 
Although this is not yet the case, we expect based on Ref.~\cite{CMS:2024irj}, that experimental improvements at the HL-LHC may allow for bin widths of at least 20\,GeV.


\subsection{Top-quark width variation}

\begin{figure}[t]
  \begin{center}  \includegraphics[width=0.49\textwidth]{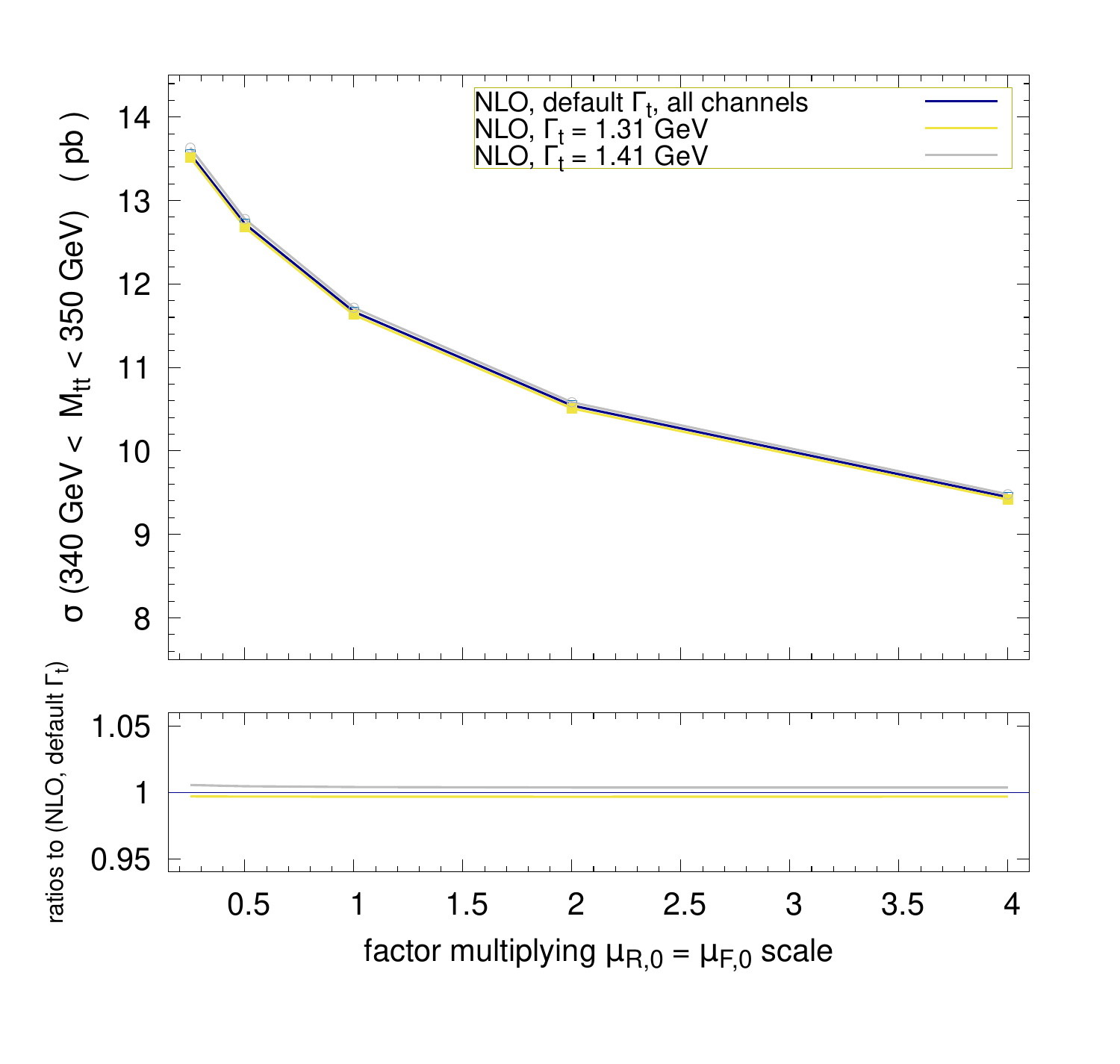}
  \caption{\label{fig:1ga} 
  Cross section in the $M_{t\bar{t}} \in [340, 350]\, \mathrm{GeV}$ bin using Eq.~\eqref{eqn:freeTrate} 
as a function of hard scale variations taken as multiples of $\mu_{R,0}=\mu_{F,0}=172.5$ GeV, varying $\Gamma_t = 1.36\pm0.05$ GeV and keeping all other parameters fixed. 
  }
 \end{center}
\end{figure}
\begin{figure}[h]
  \includegraphics[width=0.49\textwidth]{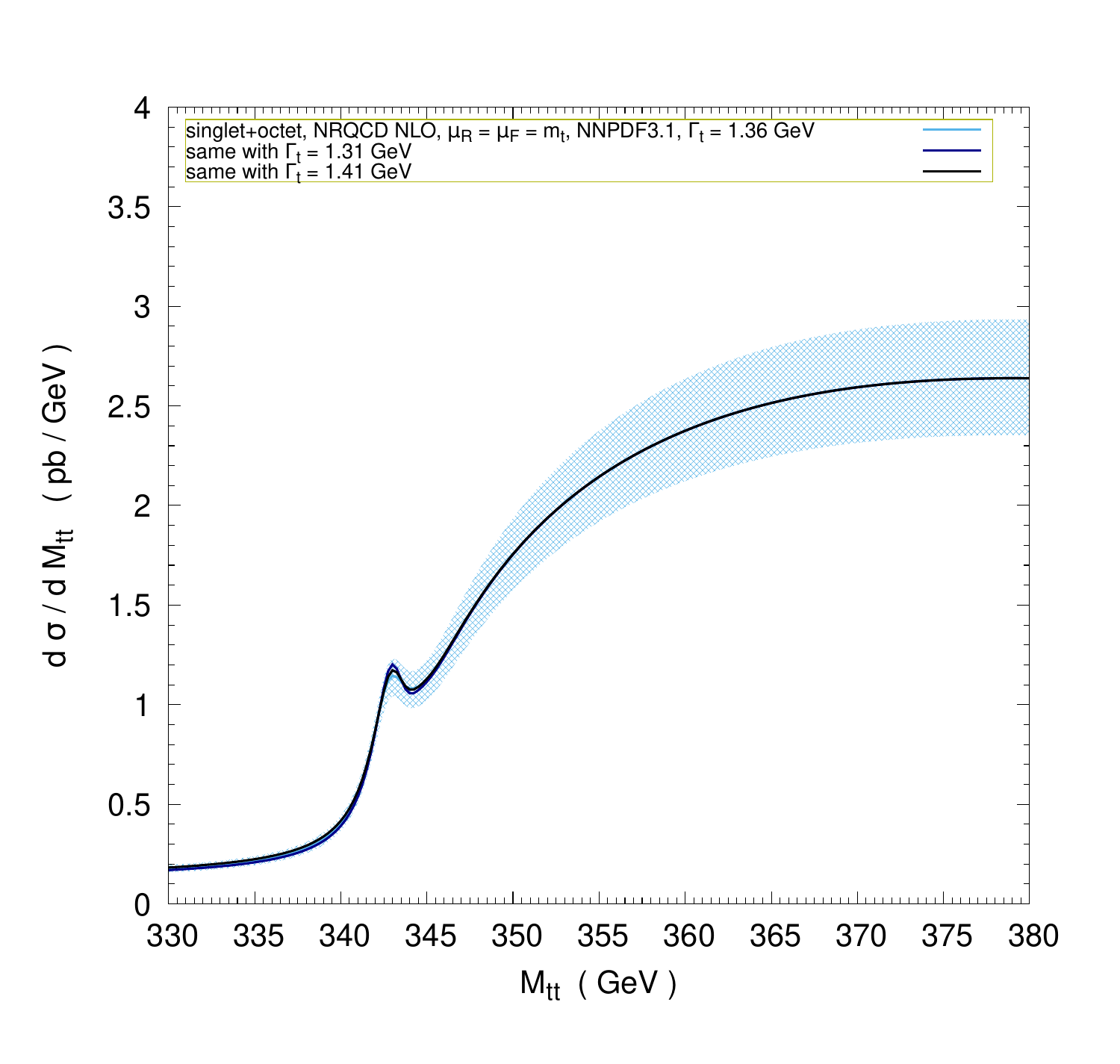}
  \includegraphics[width=0.49\textwidth]{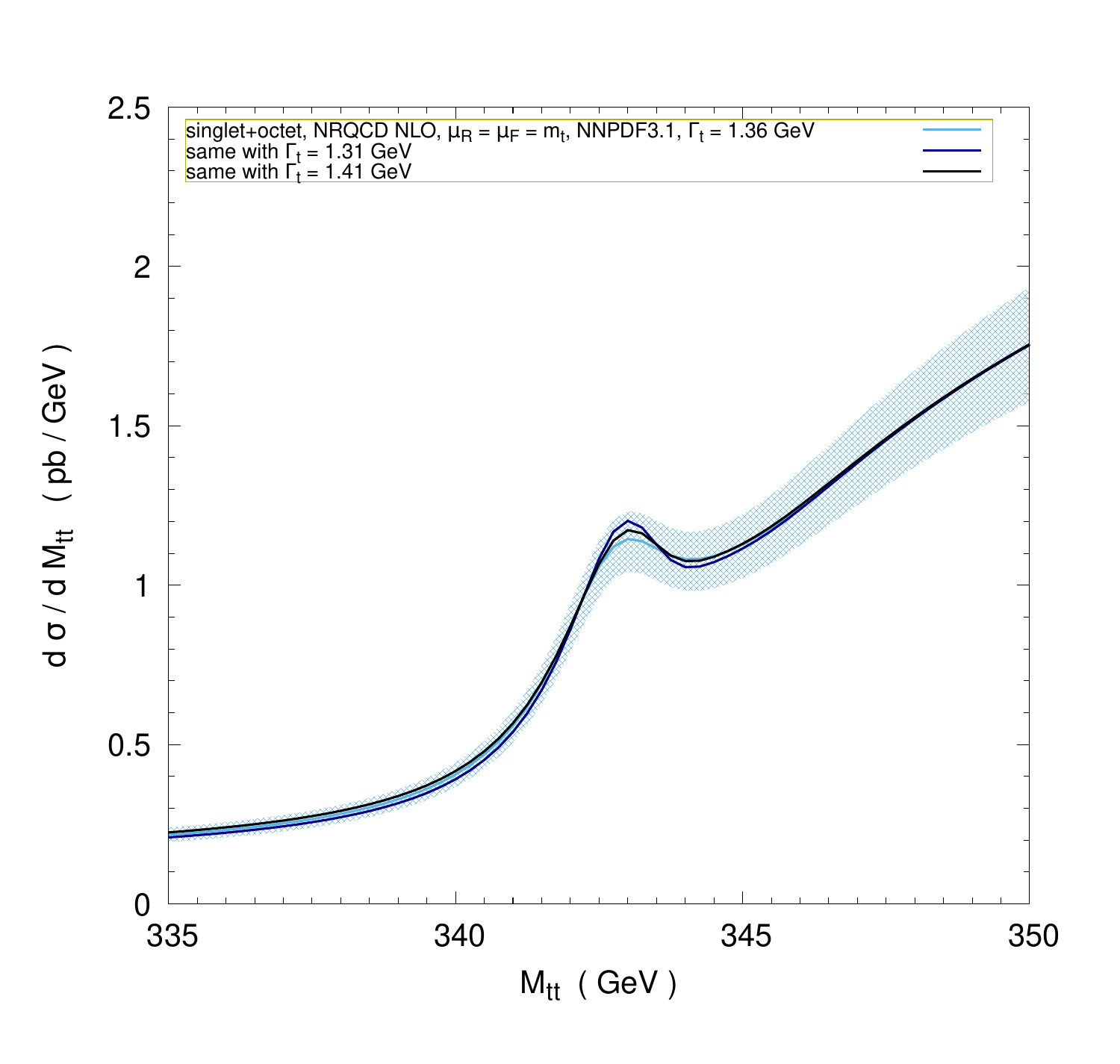}
  \caption{\label{fig:2ga} 
  Left panel: predictions for $d\sigma/dM_{t\bar{t}}$ with the central scale choice $\mu_{R,0}=\mu_{F,0}=m_t$ using the NNPDF3.1 PDF set with its default $\alpha_s(M_Z)$ value and 
  $\Gamma_t = 1.36\, \mathrm{GeV}$, compared with results for $\Gamma_t = 1.31\, \mathrm{GeV}$
  and $\Gamma_t = 1.41\, \mathrm{GeV}$. The light-blue band shows the hard scale uncertainty of the default setup. Right panel: zoom of the same comparison in the threshold region.
    }
\end{figure}

In Fig.~\ref{fig:1ga} we study the impact of varying the top-quark width $\Gamma_t$. 
In this analysis we neglect its correlation with the top-quark mass, 
even though at leading order $\Gamma_t$ scales roughly with $m_t^3$. 
The experimentally measured value is $\Gamma_t = 1.36 \pm 0.02({\rm stat}){}^{+0.14}_{-0.11}({\rm syst)}~\mathrm{GeV}$~\cite{CMS:2014mxl}, corresponding to an overall uncertainty of about $\pm 10\%$.
In comparison, theoretical predictions for $\Gamma_t$ achieve a precision better than $\pm 1\%$, see, e.g.~\cite{Chen:2023osm}. 
For our analysis, we adopt the central value $\Gamma_t = 1.36~\mathrm{GeV}$ and vary it by $\pm 0.05~\mathrm{GeV}$ as a reasonable compromise. This variation changes the predictions in the region $340 < M_{t\bar{t}} < 350\, \mathrm{GeV}$ by less than 1\%, much smaller than the effects of varying $m_t$ and also smaller than the uncertainty from $\alpha_s(M_Z)$. 
The bottom panel of Fig.~\ref{fig:1ga} also shows that this is largely independent of the choices for the scales $\mur$ and $\muf$.
The width enters our calculation through the Green's functions, but not through the functions $F_{ij\to T}$, which are computed assuming stable top-quarks. 
Within our NRQCD framework, top-quark decays are factorized in Eq.~\eqref{eqn:NRQCDformula} after the on-shell $t\bar{t}$ production stage.

Figure~\ref{fig:2ga} shows the sensitivity of the $M_{t\bar{t}}$ differential distribution to variations of the top-quark width. 
As seen in the left panel, changing $\Gamma_t$ by $\pm 0.05~\mathrm{GeV}$ around the central value has only a minor effect on the predictions, with the curves lying well within the scale uncertainty band of the default setup (light blue). 
The right panel zooms into the threshold region, where the impact of $\Gamma_t$ remains very small compared to other parametric uncertainties. 
This confirms that the present experimental uncertainty on the top-quark width plays only a subdominant role in the shape of the distribution near threshold.


\subsection{PDF variation}
\label{sec:PDFunc}

\begin{figure}[b!]
  \begin{center}
  \includegraphics[width=0.49\textwidth]{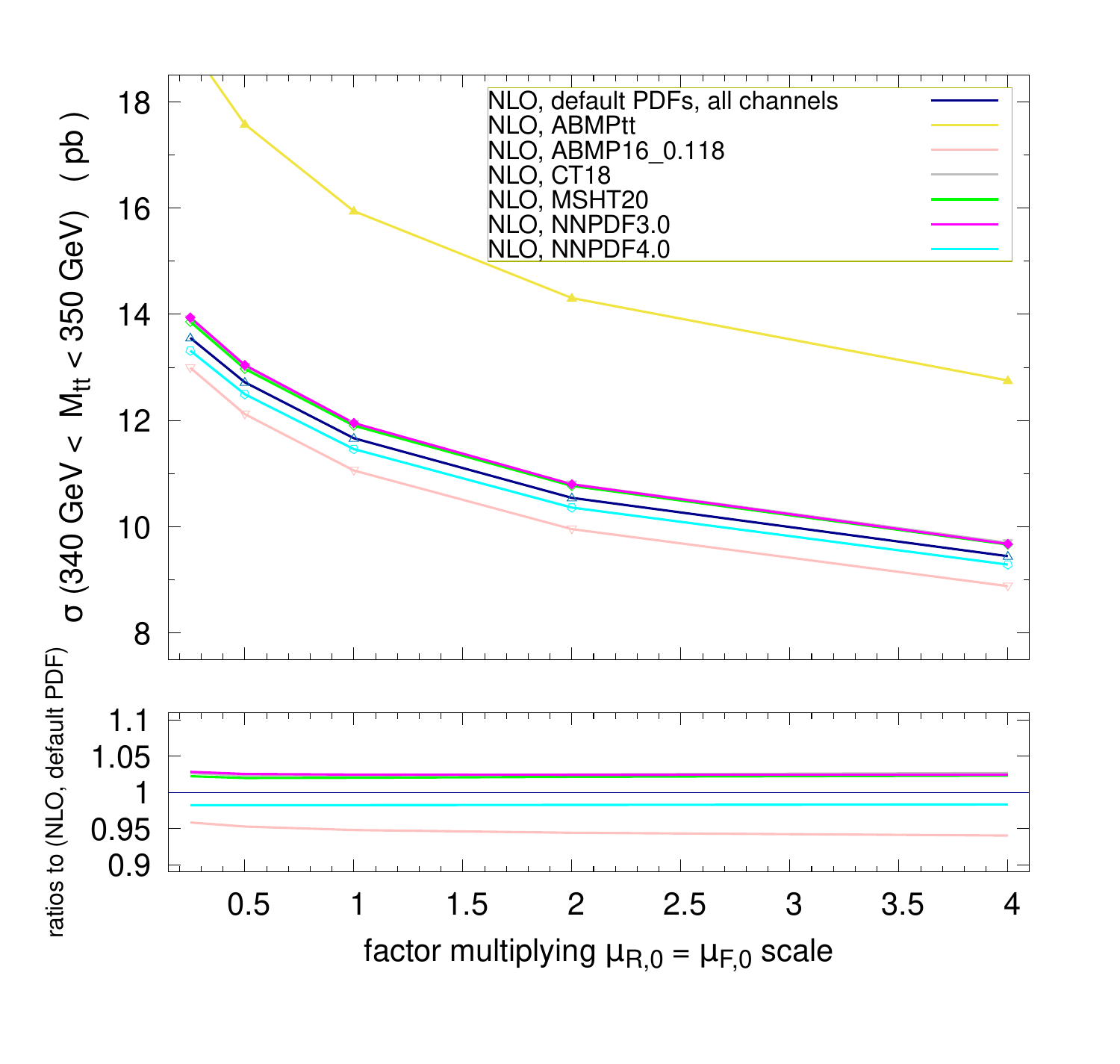}
  \caption{\label{fig:1pdf} 
Cross section in the $M_{t\bar{t}} \in [340, 350]\, \mathrm{GeV}$ bin 
as a function of hard scale variations taken as multiples of $\mu_{R,0}=\mu_{F,0}=172.5$ GeV,
using various (PDF + $\alpha_s(M_Z)$) sets at NNLO. With the PDF sets ABMP16 (with $\alpha_s(M_Z) = 0.1180$) and ABMPtt  (with $\alpha_s(M_Z) = 0.1150$) we use values of $m_t=172.5\, \mathrm{GeV}$ and $170.15\, \mathrm{GeV}$, respectively. 
  }
 \end{center}
\end{figure} 
\begin{figure}[h!]  
  \includegraphics[width=0.49\textwidth]{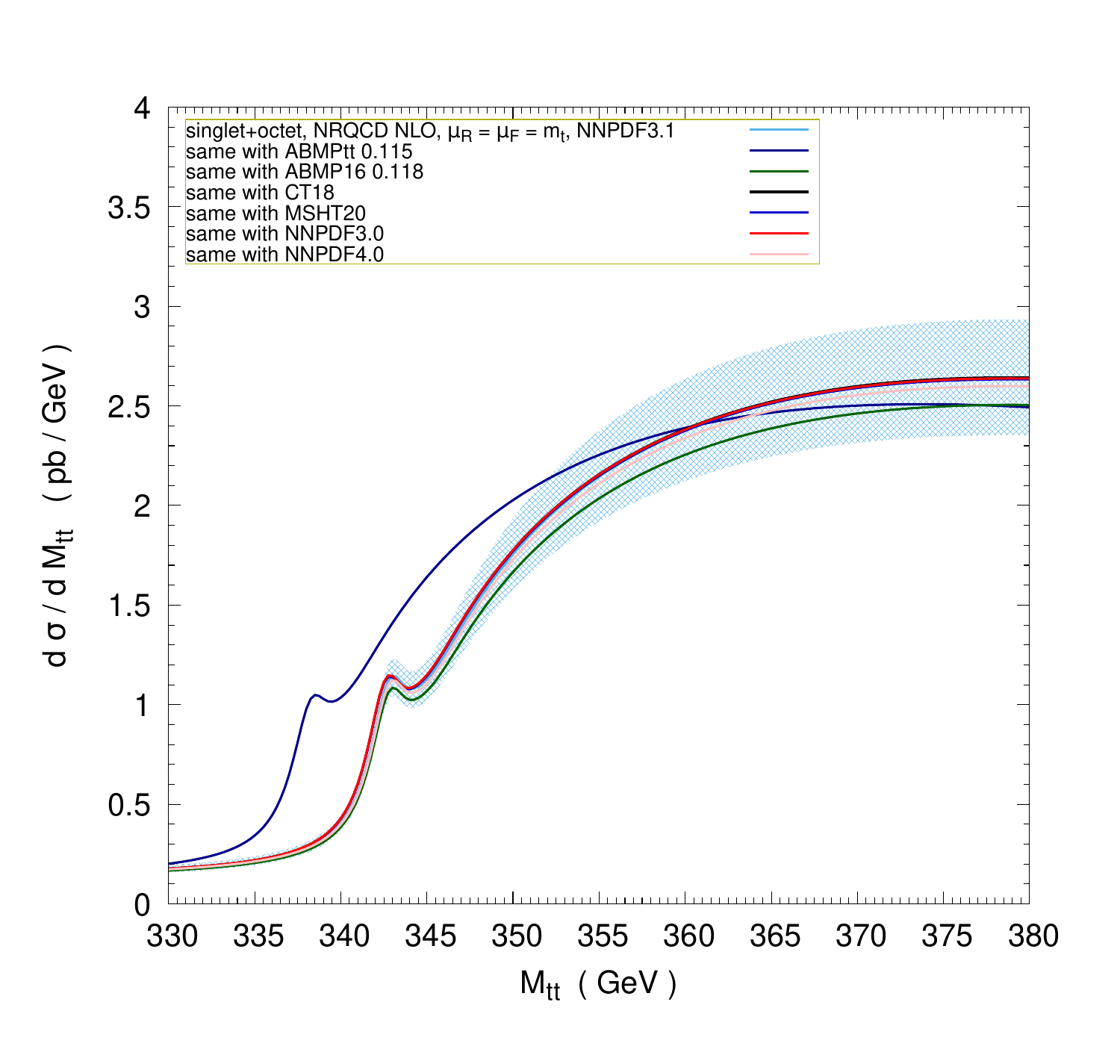}
  \includegraphics[width=0.49\textwidth]{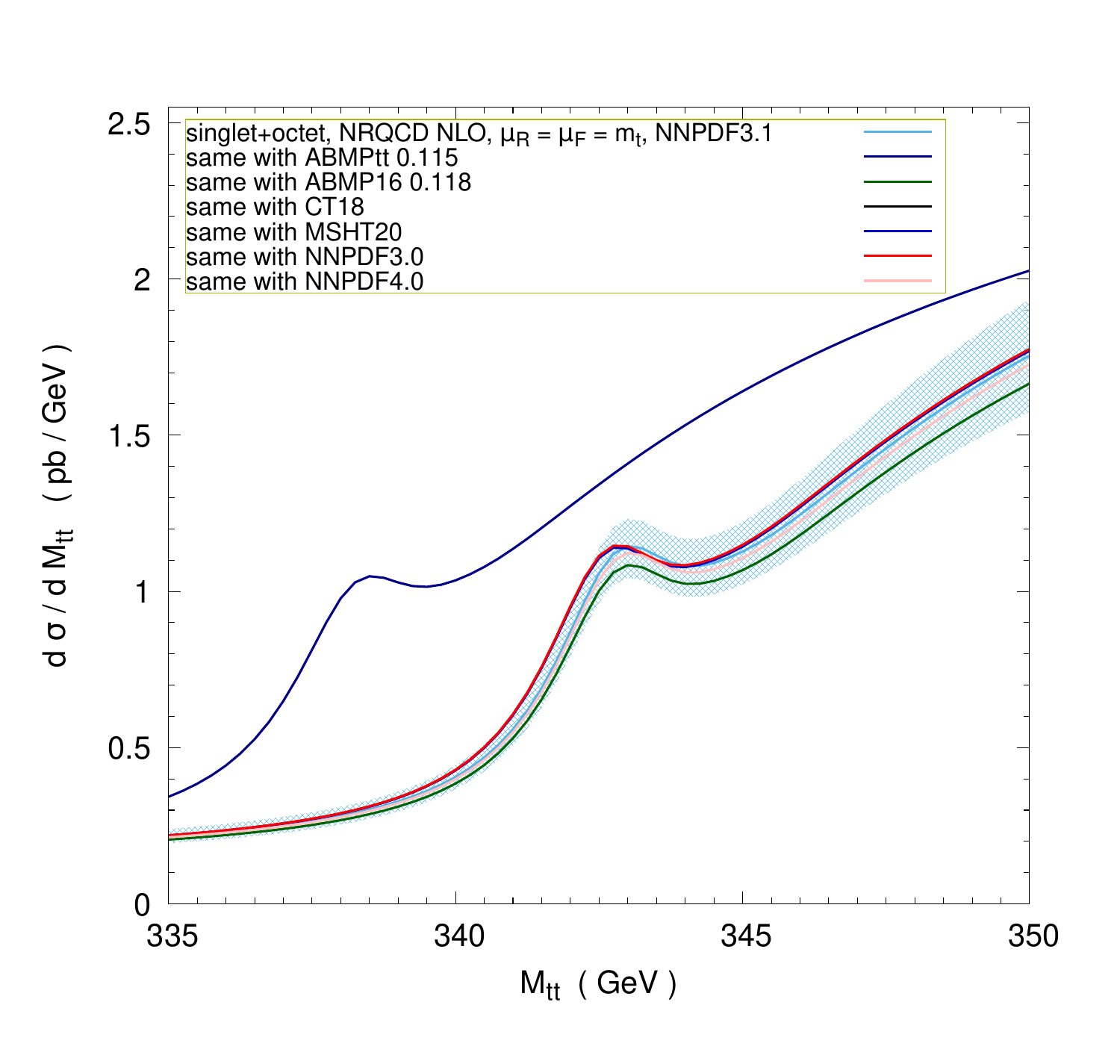}
  \caption{\label{fig:2pdf}
  Left panel: predictions for $d\sigma/dM_{t\bar{t}}$ obtained using different central PDF sets with the  central scale choice 
  $\mu_R = \mu_F = m_t$. 
  The light-blue band shows
  the hard cale uncertainty of our default setup (NNPDF3.1). Right panel: zoom of
  the same comparison in the threshold region. 
}
\end{figure}

In this section we illustrate the impact of using different PDF sets on the integrated cross section in the bin $340 < M_{t\bar{t}} < 350\, \mathrm{GeV}$. 
As shown in Fig.~\ref{fig:1pdf} (left) the predictions obtained with CT18, MSHT20, and NNPDF3.0 lie almost on top of each other, while those based on
NNPDF3.1 and NNPDF4.0 are slightly lower.
This behaviour is consistent with the variations of the predicted $M_{t\bar{t}}$ differential distribution in Fig.~\ref{fig:2pdf}.
The zoom into the threshold region in Fig.~\ref{fig:2pdf} (right) confirms that these differences remain modest. 
All these PDF sets are determined using the common input values $\alpha_s(M_Z) = 0.118$ and top-quark mass $m_t=172.5\, \mathrm{GeV}$.
Their close agreement is a consequence of the similarity of the gluon PDFs, which dominate $t\bar{t}$ production at the LHC for parton momentum fractions in the range $10^{-3} \lesssim x \lesssim 0.5$, depending on the $t\bar{t}$-pair rapidity.

In contrast, the predictions based on the ABMP16 PDF set are lower by a few percent due to differences in the PDFs, whereas those derived from ABMPtt PDFs show a shifted peak position. 
Besides differences in their gluon PDFs, this is due to the fact that the values of $\alpha_s(M_Z)$ and $m_t$ are fitted simultaneously in these analyses. 
For ABMPtt, the fitted parameters are $\alpha_s(M_Z)=0.1150$ and $m_t = 170.15\, \mathrm{GeV}$
(after converting the $\overline{\mathrm{MS}}$ mass $m_t(m_t)$ to the pole mass), 
while for ABMP16 we use the variant with $\alpha_s(M_Z)=0.1180$ and $m_t = 172.5\, \mathrm{GeV}$.

Overall, the spread among PDF sets using the same values of $m_t$ and $\alpha_s(M_Z)$
leads to only small differences in the predicted $M_{t\bar{t}}$ distribution. In all cases examined, PDF-induced variations remain significantly smaller than the scale-variation uncertainty of our default setup, indicating that PDF uncertainties play a subdominant role in the kinematic region around threshold. We estimate the PDF uncertainties from the envelope of the sets shown in Fig.~\ref{fig:2pdf} that use the same values of $\alpha_s(M_Z)$ and $m_t$ (i.e., all sets except the ABMPtt PDFs). 
This envelope is larger than the PDF uncertainty band computed with the  100 members of the NNPDF3.1\_nnlo\_as\_118 set (not shown in Fig.~\ref{fig:2pdf}), but remains well within the hard scale uncertainty band of our default setup (light-blue band). 
Therefore, it provides a more conservative estimate.


\section{Phenomenology implications for present LHC analyses}
\label{sec:impli}

Here we present the results obtained for the $t\bar{t}+X$ production cross section at the LHC within the NRQCD framework, following the approach discussed in Sec.~\ref{sec:theo}. 
In order to align our study with the analyses of the CMS~\cite{CMS:2025kzt} and ATLAS~\cite{ATLAS:2026dbe}  collaborations, we compare our predictions with those obtained from the \texttt{hvq}~\cite{Frixione:2007nw} generator within the \texttt{POWHEG-BOX}~\cite{Alioli:2010xd}, used to simulate $t\bar{t}$ pair production at NLO also accounting for top-quark decay effects. 
We provide an excess cross section together with the corresponding uncertainty based on the systematic variation of the input parameters entering our computation, as described in Sec.~\ref{sec:pred} by comparing our results with standard NLO perturbative QCD predictions.

\begin{figure}[b!]
  \includegraphics[width=0.49\textwidth]{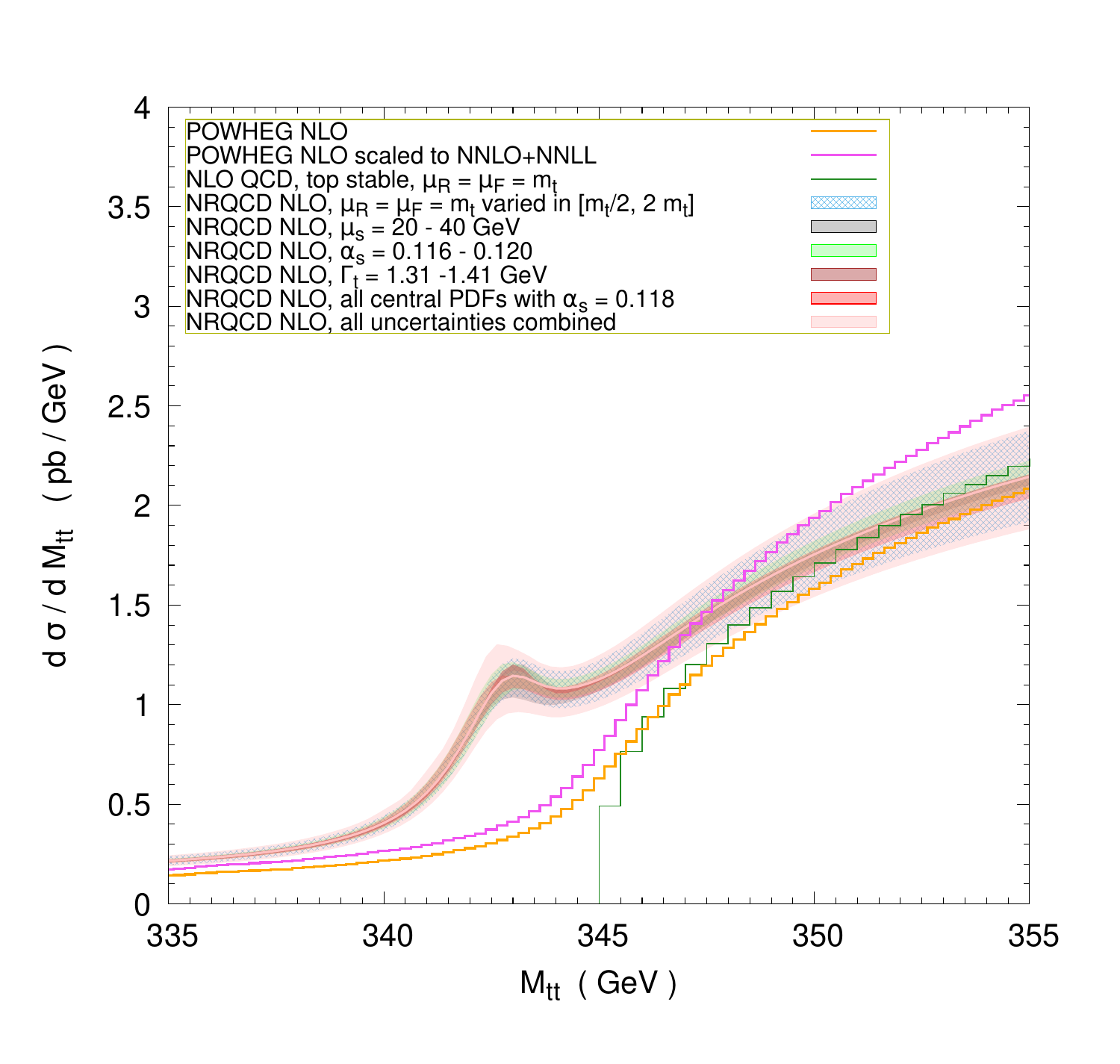}
  \includegraphics[width=0.49\textwidth]{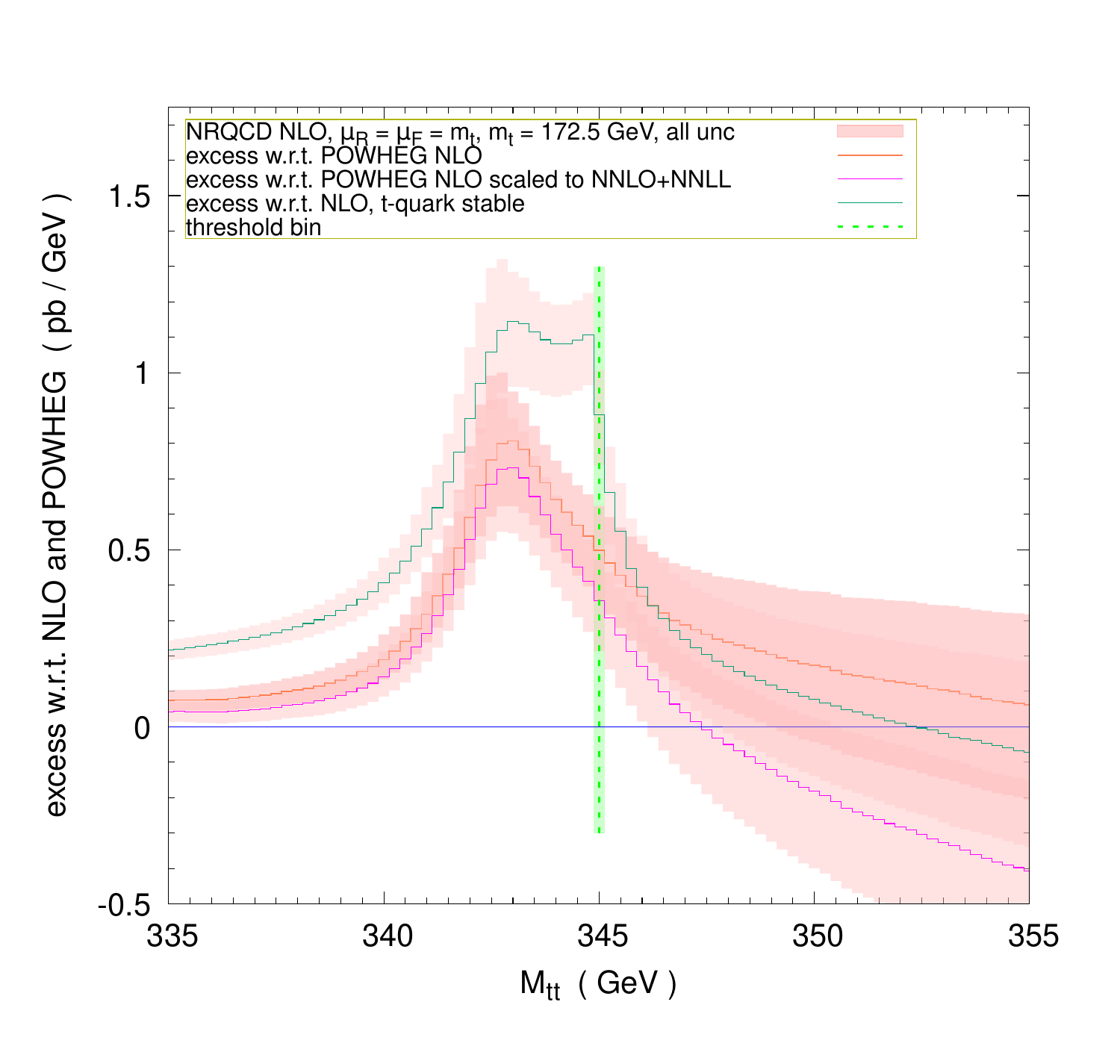}
  \caption{\label{fig:6} Left panel: NRQCD NLO predictions for $d\sigma/dM_{t\bar{t}}$ with uncertainties vs.\ fixed-order QCD NLO predictions for stable $t\bar{t}$ production and \texttt{POWHEGBOX-hvq} predictions with and without rescaling to $t\bar{t}+X$ total cross section at NNLO+NNLL, using as central hard scale $\mu_{R,0}=\mu_{F,0}=m_t$. Right panel: excess of NRQCD NLO predictions with uncertainties, using as central hard scale $\mu_{R,0}=\mu_{F,0}=m_t$, w.r.t.\ fixed-order QCD NLO ones for stable $t\bar{t}$ production and \texttt{POWHEGBOX-hvq} predictions with and without
  rescaling to $t\bar{t}+X$ total cross section at NNLO+NNLL. The vertical green dashed line highlights the threshold region.
  }
\end{figure}
In the left panel of Fig.~\ref{fig:6} we show the results for the differential distribution $d\sigma/dM_{t\bar{t}}$ in the threshold region, within the NRQCD framework, for a central scale choice $\mu_{R,0}=\mu_{F,0}=m_t=172.5$~GeV. We also include fixed-order NLO predictions for the production of stable $t\bar{t}$ pairs, as well as predictions accounting for top-quark decays. 
Following the approach adopted in the CMS analysis~\cite{CMS:2025kzt}, the NLO distribution $d\sigma/dM_{t\bar{t}}$ for top-quark production and decay has also been rescaled to reproduce the total cross section for $t\bar{t}$ production at NNLO with additional threshold resummation up to next-to-next-to-leading logarithmic (NNLL) accuracy~\cite{Czakon:2011xx}. 
In the NRQCD regime, the inclusion of quasi-bound state effects leads to a significant enhancement of the cross section in the threshold region compared to standard perturbative QCD predictions. 
This effect is also reported in the right panel of Fig.~\ref{fig:6}, showing the NRQCD excess as a function of the invariant mass $M_{t\bar{t}}$.
The pink uncertainty bands in both panels in Fig.~\ref{fig:6} 
refer to the sum in quadrature of the uncertainties associated with the variation of the PDF set, $\mu_s,~\mu_{R,0},~\mu_{F,0},~\as,~\Gamma_t$ and resulting in a total uncertainty of approximately $\pm12\%$ for the NRQCD prediction in bin $M_{t\bar{t}}\in[340,350]$ GeV around the peak region.
The uncertainty due to the PDF sets variation is computed from the envelope of predictions from six NNLO PDF sets, as discussed in Sec.~\ref{sec:PDFunc}. 
The uncertainty related to the $\Gamma_t$ variation is assumed to be independent of the $m_t$ value, which has been chosen consistently with the adopted PDF sets. 
The right panel of Fig.~\ref{fig:6} clearly shows the validity region for NRQCD predictions, with the uncertainty band becoming larger as long as high invariant mass regions are probed.
This phase space region is well described by predictions based on perturbative QCD, which are more sensitive to hard-scale variation and to missing higher orders in a fixed-order expansion than the threshold region, which is dominated by the uncertainties stemming from $\mu_s$ variation, and from higher-order effects in the NRQCD expansion. 

From the pink curve in the right panel of Fig.~\ref{fig:6} we observe that the inclusion of NLO results rescaled to the NNLO+NNLL total cross section reduces the size of the NRQCD excess over the entire invariant mass region under investigation, which even becomes negative for $M_{t\bar{t}}\sim 348~\rm{GeV}$. Since our predictions are solely based on the implementation of Eq.~\eqref{eqn:NRQCDformula}, in the remainder of our analysis we will no longer include any rescaling on the cross section.

Based on these considerations, 
we provide a list of the ingredients needed to determine the  cross section enhancement due to quasi-bound state effects:
\begin{enumerate}
\item Compute  base-line predictions based on perturbative QCD expansion
\begin{itemize}
\item Use a parton-level event generator to simulate $t{\bar t}$ production, also accounting for top-quark decay effects
     (e.g. the \texttt{hvq} generator within \texttt{POWHEG-BOX} at NLO in the NWA).  
\item These predictions serve as the \emph{``pQCD''} baseline $\sigma^{\rm{pQCD}}$.
\end{itemize}
\item Compute the NRQCD prediction  
\begin{itemize}
\item Ensure that the calculation uses the same $M_{t{\bar t}}$ binning and kinematic settings adopted for $\sigma^{\rm{pQCD}}$.
\item 
Evaluate the cross section $\sigma^{\rm{NRQCD}}$ for $t\bar{t}+X$ production, accounting for the resummation of quasi-bound state effects within the NRQCD framework
(see Eq.~\eqref{eqn:NRQCDformula} for the NLO NRQCD expression).  
\end{itemize}
\item Determine the excess  
\begin{itemize}
\item Ensure that the perturbative orders are consistent on both sides (neither ad-hoc rescaling, nor reweighting etc.). 
\item Compute the cross section enhancement $\sigma^{\rm{excess}}$ as the difference between $\sigma^{\rm{NRQCD}}$ and $\sigma^{\rm{pQCD}}$.
\end{itemize}
\end{enumerate}

We are now able to isolate the cross section enhancement by integrating numerically Eq.~\eqref{eqn:QCDmasterformula} within the peak region corresponding to $M_{t\bar{t}}\in[340,350]$ GeV.
This leads to
\begin{equation}
\label{eqn:xs_NRQCD}
\sigma^{\rm{NRQCD}}=
\int_{340~\rm{GeV}}^{350~\rm{GeV}}dM_{t\bar{t}}
\frac{d\sigma_{P_1 P_2\to T}}{d M_{t\bar{t}}}=
11.67^{+1.43}_{-1.47}~\rm{pb}\,.
\end{equation}
The breakdown into the individual uncertainites due to variations of the PDF set, $\mu_s,~\mu_{R,0},~\mu_{F,0},~\as$ and $\Gamma_t$ (added in quadrature, keeping $\mu_{R,0}=\mu_{F,0}$) is as follows:
\begin{equation}
\label{eqn:xs_NRQCD-unc}
\left(\Delta \sigma^{\rm{NRQCD}}/\rm{pb}\right)^2 =
\left({}^{+1.43}_{-1.47}\right)^2 
= 
\left({}^{+1.05}_{-1.12}\right)_{\mu_{R,0},\,\mu_{F,0}}^2 +
\left({}^{+0.79}_{-0.55}\right)_{\mu_s}^2 +
\left({}^{+0.49}_{-0.47}\right)_{\as}^2 +
\left({}^{+0.30}_{-0.61}\right)_{\text{PDF}}^2 +
\left({}^{+0.05}_{-0.04}\right)_{\Gamma_t}^2 
\, .
\end{equation}
Next we subtract $\sigma^{\rm{pQCD}}$
provided by the \texttt{POWHEG-hvq} generator, also accounting for top-quark decay, and obtain
\begin{equation}
    \label{eqn:excess_mu1}
    \sigma^{\rm{excess}}(340~{\rm{GeV}}\le M_{t\bar{t}}\le 350~{\rm{GeV}})=
    4.15^{+1.43}_{-1.47}~\rm{pb}.
\end{equation}
This result strongly depends on the selected integration range,
as described in Tab.~\ref{tab:excessmu1}, in which we report $\sigma^{\rm excess}$ for a stable $t\bar{t}$ pair, and with the inclusion of top-quark decay effects, for a number of $M_{t\bar{t}}$ bins. 
A rigorous choice of the integration range requires the formulation of a matching prescription to combine the NRQCD results, prominent in the threshold region, with the usual fixed-order perturbative QCD predictions, which are dominant in the continuum. We leave the formulation of a matching for a future and dedicated work.

\begin{table}[b!]
    \centering
{\footnotesize
    \begin{tabular}{|c|c|c|c|c|}  
    \hline
        $M_{t\bar{t}}$ range 
        & $\sigma_{\rm{NLO}}^{\rm{NRQCD}}$ 
        & $\Delta_{\sigma}$ 
        & $\sigma^{\rm{excess}}$  w.r.t. 
        & $\sigma^{\rm{excess}}$ w.r.t. 
        \\
        (GeV) &  (pb)
& (pb) 
& NLO  $t\bar{t}$ stable (pb) & hvq + $t\bar{t}$ decay (pb)\\ 
        \hline
340 - 345  &      4.42  & $^{+ 0.62}_{ - 0.60}$&     4.42  &     2.73     \\
\hline
340 - 346  &      5.61  & $^{+ 0.75}_{ - 0.74}$&     4.98  &     3.16     \\
\hline
340 - 347  &      6.92  & $^{+ 0.90}_{ - 0.90}$&     5.28  &     3.48     \\
\hline
340 - 348  &      8.38  & $^{+ 1.06}_{ - 1.08}$&     5.49  &     3.74     \\
\hline
340 - 349  &      9.97  & $^{+ 1.24}_{ - 1.27}$&     5.63  &     3.97     \\
\hline
340 - 350  &     11.67  & $^{+ 1.43}_{ - 1.47}$&     5.73  &     4.15     \\
\hline
340 - 351  &     13.47  & $^{+ 1.64}_{ - 1.69}$&     5.79  &     4.31     \\
\hline
340 - 352  &     15.37  & $^{+ 1.85}_{ - 1.92}$&     5.81  &     4.44     \\
\hline
340 - 353  &     17.34  & $^{+ 2.08}_{ - 2.15}$&     5.81  &     4.56     \\
\hline
340 - 354  &     19.39  & $^{+ 2.31}_{ - 2.40}$&     5.77  &     4.65     \\
\hline
340 - 355  &     21.50  & $^{+ 2.55}_{ - 2.66}$&     5.71  &     4.72     \\
        
      \hline
    \end{tabular}
}\caption{\label{tab:excessmu1} The cross section enhancement due to NRQCD effects for $\mu_{R,0}=\mu_{F,0}=m_t=172.5$ GeV, for several invariant mass regions, with respect to NLO predictions for on-shell $t\bar{t}$ production (fourth column), and for $t\bar{t}$ production and decay (last column).
}
\end{table}

\begin{figure}
  \includegraphics[width=0.49\textwidth]{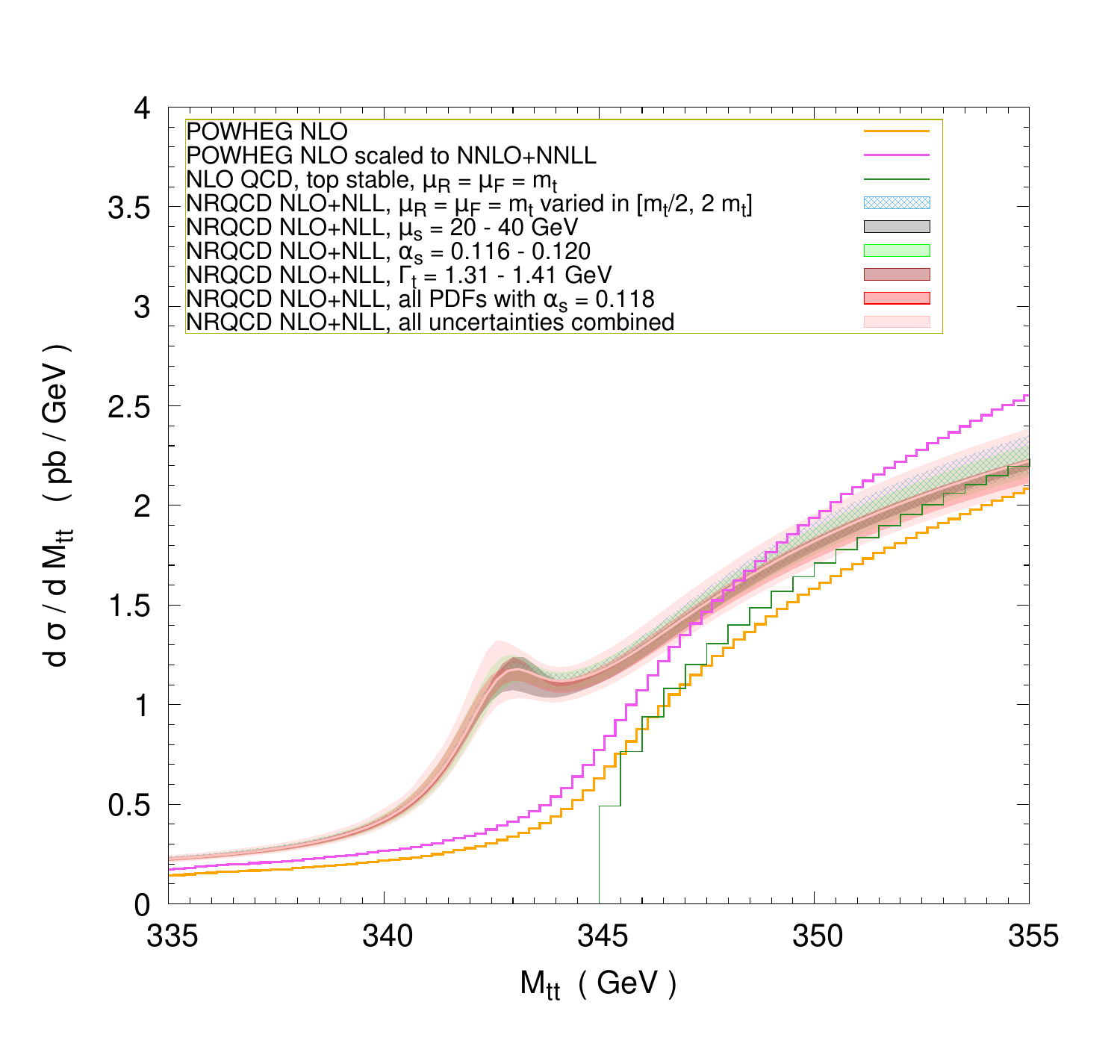}
  \includegraphics[width=0.49\textwidth]{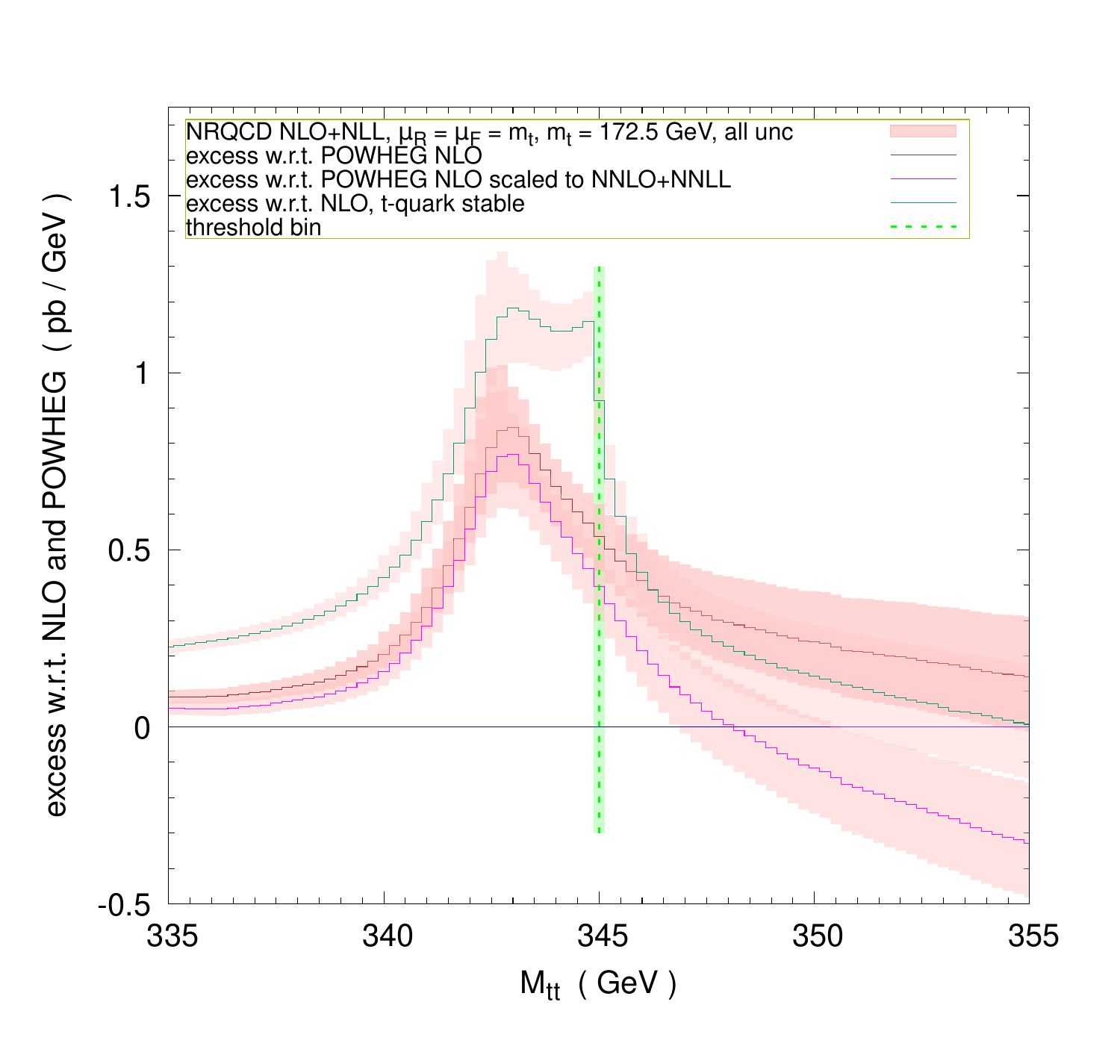}
  \caption{\label{fig:8} 
    Same as Fig.~\ref{fig:6}, with the inclusion of NLL soft-gluons threshold resummation.
    }
\end{figure}
In Fig.~\ref{fig:8} we display the same curves as in Fig.~\ref{fig:6}, showing how the NLL resummation of threshold-enhanced logarithms arising in the hard function $F_{ij}$ impacts the NRQCD predictions, reducing the uncertainty band across the whole $M_{t\bar{t}}$ region under investigation. 
As in Fig.~\ref{fig:6}, the right panel of Fig.~\ref{fig:8} shows that adopting NLO predictions rescaled to the NNLO+NNLL total cross section leads to a reduced NRQCD enhancement throughout the whole phase space region considered.
Furthermore, the resummation of soft-gluon logarithms also impacts over the normalisation of the NRQCD distribution, enhancing $\sigma^{\rm{excess}}$ up to $\sim10\%$, for $M_{t\bar{t}}$ in $[340,350]$~GeV, with the enhancement strongly depending on the selected $M_{t\bar{t}}$ bins, as reported in Tab.~\ref{tab:excessmu1res}. 
The numerical value of $\sigma^{\rm{excess}}$ also depends on the adopted value for the top-quark mass $m_t$, which directly affects the position of the \emph{toponium}-like peak, located at $M_{t\bar{t}}\simeq 2m_t$, as reported in Fig.~\ref{fig:2mt}. 
Further details on individual uncertainties, in analogy to Eq.~\eqref{eqn:xs_NRQCD-unc} are given in App.~\ref{app:XSunc-xtd} and~\ref{app:further_tables}.
\begin{figure}[t]
    \begin{center}
\includegraphics[width=0.49\textwidth]{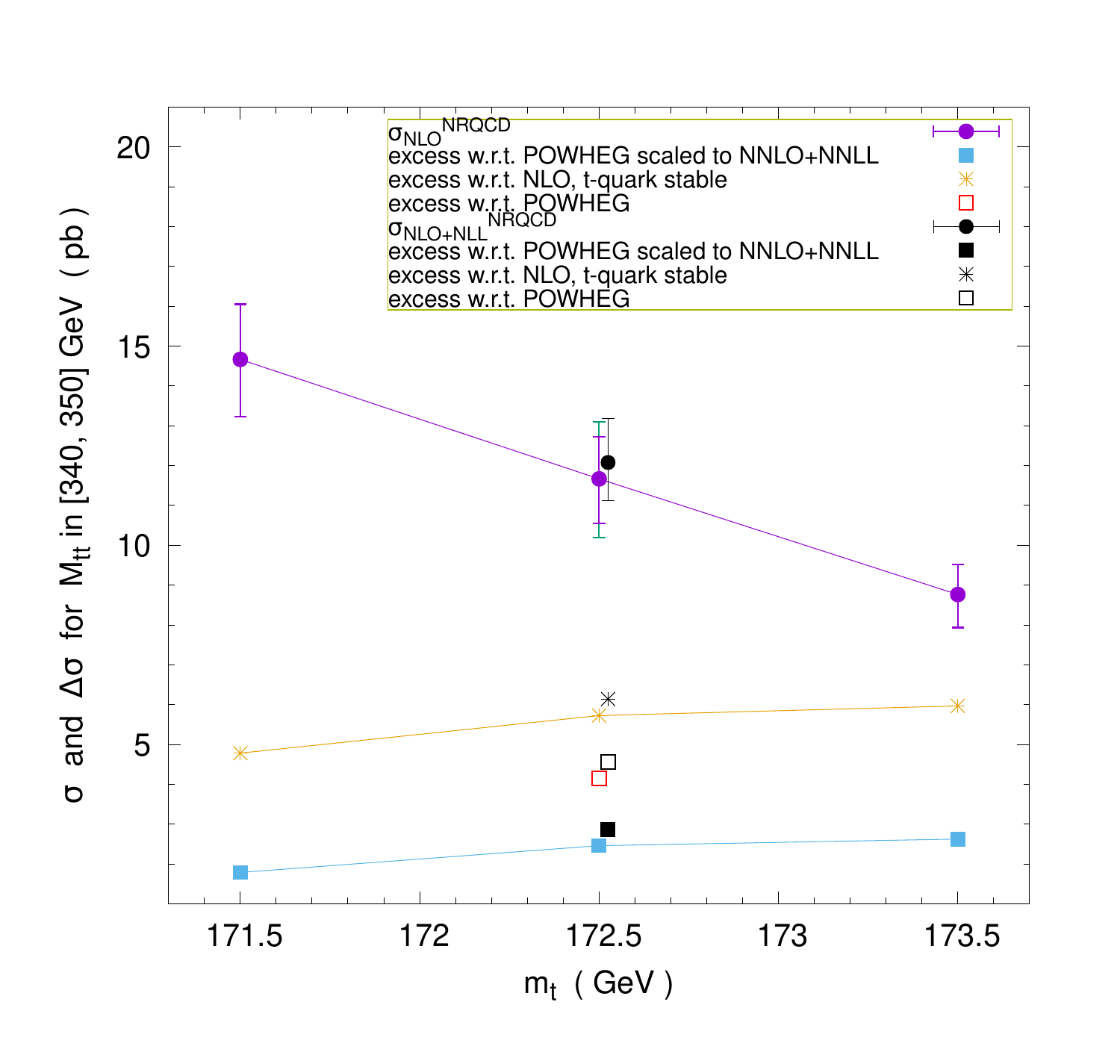}
\includegraphics[width=0.49\textwidth]{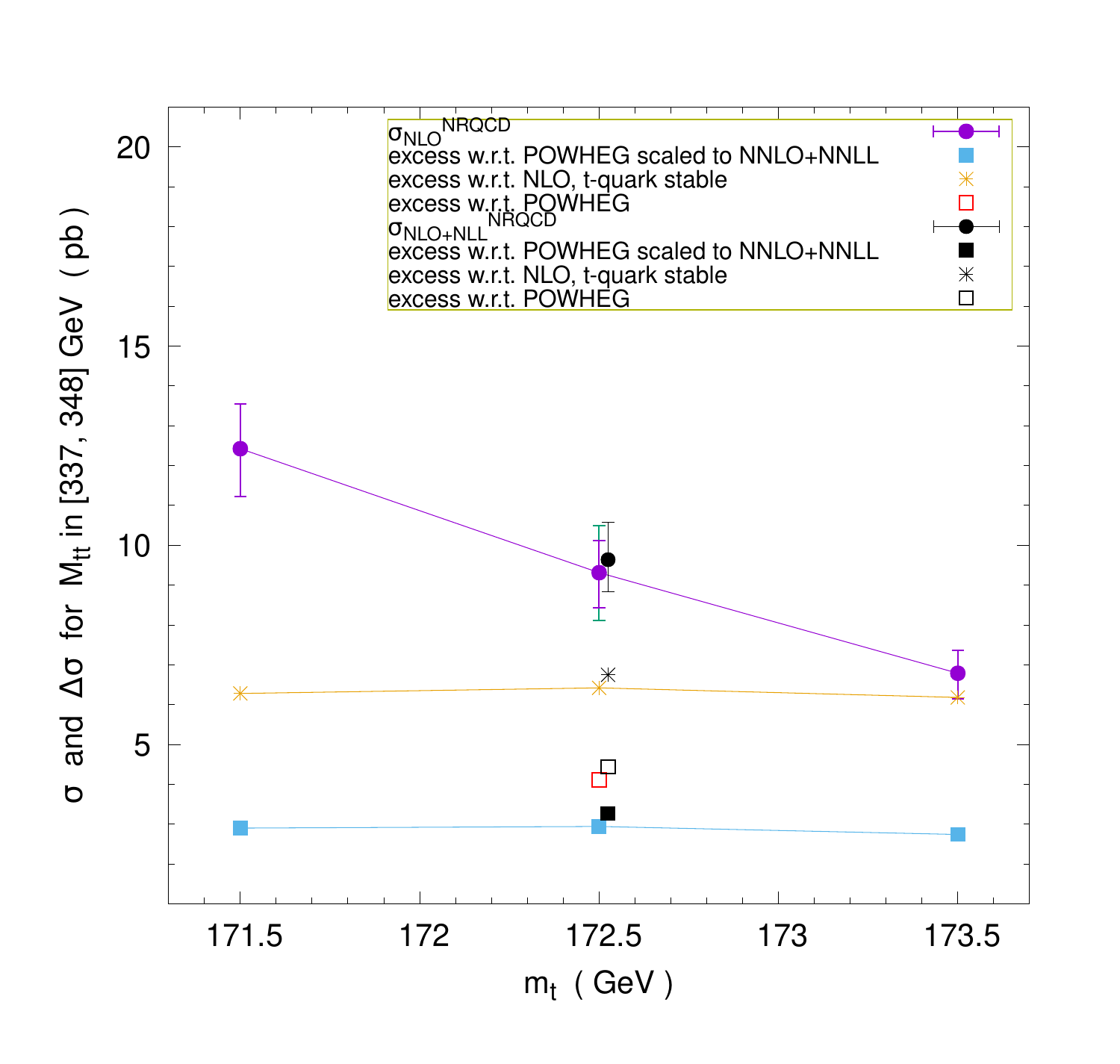}
    \caption{Predictions for $\sigma^{\rm{NRQCD}}$ (violet), $\sigma^{\rm{excess}}$ with respect to NLO perturbative QCD results for a stable $t\bar{t}$ pair (orange), and $\sigma^{\rm{excess}}$  with respect to the NLO cross section for top-quark production and decay from \texttt{POWHEG-hvq}, rescaled to the NNLO+NNLL $t\bar{t}+X$ total cross section (blue) for $M_{t\bar{t}} \in [340,350]$ GeV (left panel) and in $[337,348]$ GeV (right panel). 
    The purple error bars are associated to hard-scale variations, while the green error bars were obtained considering the envelope of the uncertainties associated with the variations of all the input parameters.
    The red marker refers to $\sigma^{\rm{excess}}$ with respect to \texttt{POWHEG-hvq} predictions for $m_t=172.5$ GeV, without any rescaling factor. Each point is associated with a specific value for the top-quark mass, varied in $m_t=172.5\pm1$ GeV. The black markers (from the top to the bottom of the figure) show the results obtained with $m_t=172.5$ GeV for $\sigma^{\rm{NRQCD}}\,,\sigma^{\rm{excess}}$ with respect to NLO stable $t\bar{t}$ production, $\sigma^{\rm{excess}}$ with respect to \texttt{POWHEG-hvq} predictions, and $\sigma^{\rm{excess}}$ with respect to \texttt{POWHEG-hvq} rescaled to the NNLO+NNLL $t\bar{t}+X$ total cross section, when NLL soft-gluons threshold resummation is considered. The associated black error bars account for all the uncertainties sources.}
    \label{fig:plot_mt172.5}
        \end{center}
\end{figure}
\begin{figure}[t!]
    \begin{center}
\includegraphics[width=0.49\linewidth]{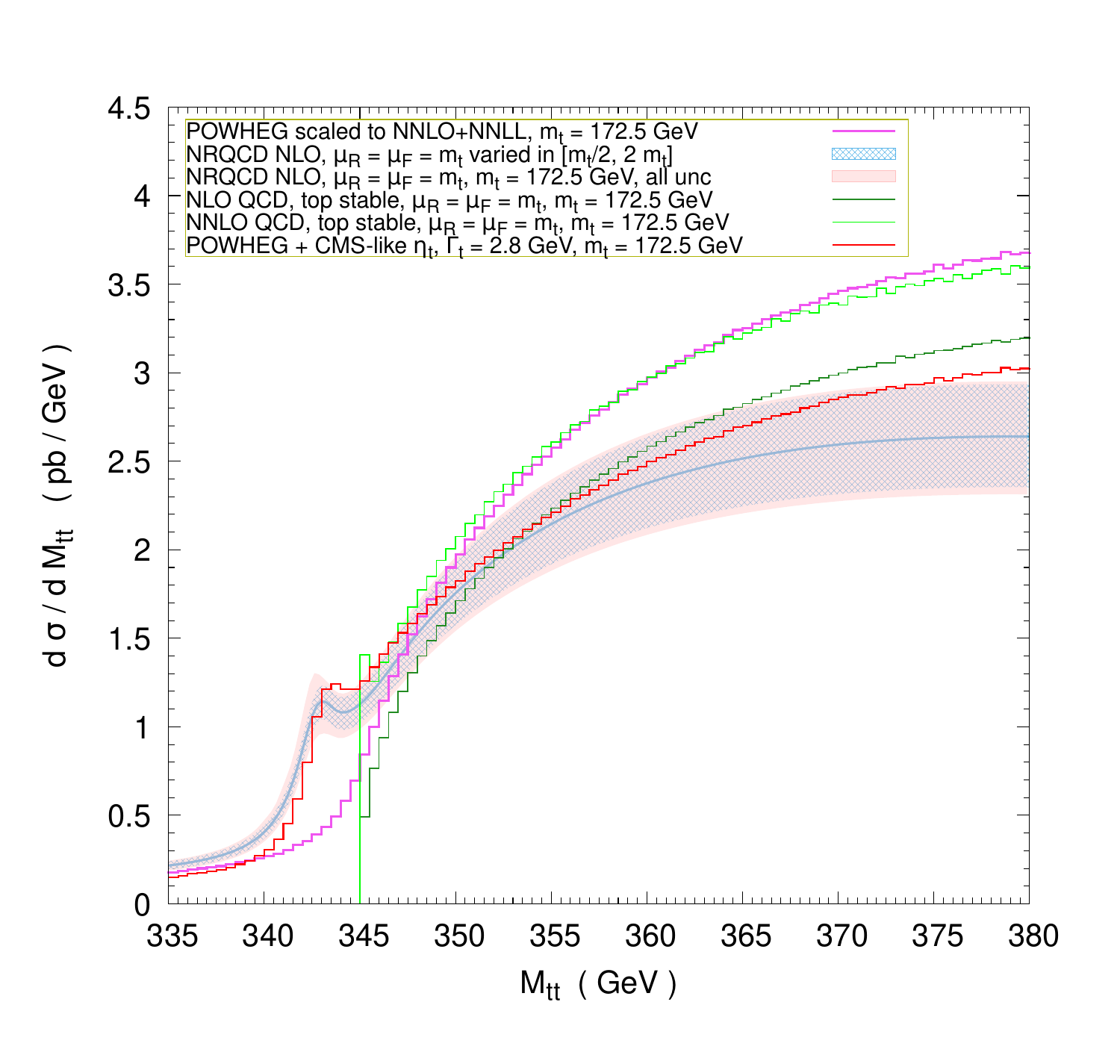}
    \caption{NLO NRQCD (light-blue and pink bands) predictions for $M_{t\bar{t}}$ distribution, compared to NLO (dark-green curve) and NNLO (light-green curve) fixed-order perturbative QCD results for stable $t\bar{t}$ pair production. The NLO differential distribution accounting for top-quark pair production and decay, rescaled to the NNLO+NNLL total cross section for $t\bar{t}+X$, is reported in purple. The distribution in red shows the NLO perturbative QCD distribution for $M_{t\bar{t}}$, including the contribution from the pseudoscalar resonance $\eta_t$, in the threshold region.}
    \label{fig:NRQCDvsCMS}
    \end{center}
\end{figure}

\begin{table}[t!]
    \centering
{\footnotesize
    \begin{tabular}{|c|c|c|c|c|c|c|c|}  
    \hline
        $M_{t\bar{t}}$ range 
        & $\sigma_{\rm{NLO+NLL}}^{\rm{NRQCD}}$ 
        & $\Delta_\sigma$ 
        & $\sigma^{\rm{excess}}$  w.r.t. 
        & $\sigma^{\rm{excess}}$ w.r.t. 
        \\
        (GeV) &  (pb)
& (pb) 
& NLO  $t\bar{t}$ stable (pb) & hvq + $t\bar{t}$ decay (pb)\\ 
        \hline
        340 - 345  &      4.57  &$^{+ 0.53}_{ -0.44}$ &     4.57  &     2.88      \\
        \hline
340 - 346  &      5.80  &$^{+ 0.62}_{ -0.53}$ &     5.17  &     3.35      \\
\hline
340 - 347  &      7.16  &$^{+ 0.73}_{ -0.63}$ &     5.52  &     3.72      \\
\hline
340 - 348  &      8.67  &$^{+ 0.85}_{ -0.74}$ &     5.78  &     4.03      \\
\hline
340 - 349  &     10.31  &$^{+ 0.97}_{ -0.85}$ &     5.98  &     4.31      \\
\hline
340 - 350  &     12.08  &$^{+ 1.10}_{ -0.96}$ &     6.14  &     4.56      \\
\hline
340 - 351  &     13.95  &$^{+ 1.23}_{ -1.08}$ &     6.26  &     4.79 \\
\hline
340 - 352  &     15.92  &$^{+ 1.37}_{ -1.21}$ &     6.36  &     4.99      \\
\hline
340 - 353  &     17.97  &$^{+ 1.52}_{ -1.34}$ &     6.43  &     5.18      \\
\hline
340 - 354  &     20.09  &$^{+ 1.66}_{ -1.47}$ &     6.47  &     5.35      \\
\hline
340 - 355  &     22.28  &$^{+ 1.82}_{ -1.61}$ &     6.49  &     5.50      \\
\hline

      \hline
    \end{tabular}
}
    \caption{\label{tab:excessmu1res} Same as for Tab.~\ref{tab:excessmu1}, but
for NLO+NLL NRQCD predictions 
}    
\end{table}

In Fig.~\ref{fig:plot_mt172.5} we report, for two different integration ranges, different values for $\sigma^{\rm{excess}}$, each obtained through the variation of the central value for the top-quark mass $m_t=172.5$ GeV by 1~GeV up and down.
When considering $M_{t\bar{t}} \in [340,350]$ GeV we observe how the 1 GeV variation on the top-quark mass impacts the value of $\sigma^{\rm{excess}}$ of approximately $\pm10\%$, with a further positive $10\%$ provided from the soft-gluon threshold resummation, that we only included for the central mass value $m_t=172.5$ GeV. On the other hand, 
we observe that the choice of the $[337,348]$ GeV integration range minimizes the dependence of $\sigma^{\rm{excess}}$ on $m_t$ variations.

We conclude this section reporting in Fig.~\ref{fig:NRQCDvsCMS} our predictions for NLO NRQCD cross section, compared to the results from the theoretical model~\cite{Fuks:2021xje} adopted in the CMS analysis~\cite{CMS:2025kzt}. This consists in the introduction of a pseudoscalar resonance $\eta_t$~\cite{Maltoni:2024tul} which decays into a $t\bar{t}$ pair, and whose mass and decay width are fitted in the threshold region for $t\bar{t}$ production (red curve in Fig.~\ref{fig:NRQCDvsCMS}, cf.~\cite{ATLAS:2021jgj,CMS:2025kzt}). 
Although the introduction of $\eta_t$ well describes the NRQCD dynamic relevant in the threshold region, we emphasize that these results are not as solid as the NRQCD predictions presented in this work, which also allow for a more reliable estimate of the uncertainty affecting the cross section enhancement due to quasi-bound state effects.


\section{Conclusions}
\label{sec:conclu}

We have conducted a comprehensive analysis of the uncertainties affecting the $M_{t\bar{t}}$ distribution in the region close to and below the production threshold based on the NRQCD formalism, which we have briefly summarized in Sec.~\ref{sec:theo}.
In particular, we have examined the impact of variations in the hard and soft scales, as well as uncertainties associated with the PDFs, $\alpha_s$, $m_t$ and $\Gamma_t$. 
Our analysis in Sec.~\ref{sec:pred} has been performed for Run 2 conditions at the LHC with a center-of-mass energy of $\sqrt{S} = 13\, \mathrm{TeV}$, which corresponds to the published experimental studies of toponium by the ATLAS and CMS collaborations.
To our knowledge, this is the first study that systematically investigates all these sources of uncertainty for toponium production at the LHC. 

These uncertainty studies help to quantify the excess cross section on top of the fixed-order plus parton shower predictions used by the experiments. 
Overall, when comparing our NRQCD predictions with the combined \texttt{POWHEG-BOX} and \texttt{hvq} predictions, using the same scale choice $\mu_R=\mu_F=m_t=172.5\, \mathrm{GeV}$, we find an excess of approximately $\sigma^{\rm{excess}} = 4.15^{+1.43}_{-1.47}\, \mathrm{pb}$ in the 
$M_{t\bar{t}}$ bin $[340, 350]~\mathrm{GeV}$, which is roughly centered around the peak region. 
This excess originates from all-order Coulomb corrections and bound-state formation in the color-singlet channel, which are included in the NRQCD calculation but are not accounted for in the set-up using \texttt{hvq} and \texttt{POWHEG-BOX}.
We emphasize that color-octet configurations are also included in our NRQCD framework and contribute not only above but also below threshold. 
Our NRQCD computation includes all $S$-wave states.
The theoretical uncertainty associated with our result for a fixed value of $m_t$, is driven by the hard and soft scale variations. 
For example, in the $M_{t\bar{t}}$ bin $[340, 350]~\mathrm{GeV}$ it amounts to 12\% for the NLO NRQCD prediction (and 8\% 
for the NLO+NLL NRQCD one). 
Improvements on this point would require extending our NRQCD analysis to NNLO accuracy.

The numerical value of the excess 
extracted with the procedure described above is somewhat smaller than those reported in recent LHC analyses by the ATLAS and CMS collaborations, although it should be noted that these measurements employ larger $M_{t\bar{t}}$ bins. 
A more detailed comparison would require matching our NRQCD calculation to fixed-order perturbative predictions, following Sec.~\ref{sec:impli}. 
We have found that the dominant source of uncertainty in the threshold region is the variation of the top-quark mass, which induces changes of several tens of percent in the predicted distribution. 
Consequently, a more refined comparison to the excess cross sections reported by ATLAS and CMS will need a careful treatment of the uncertainties associated with $m_t$, both from the theoretical side, for example the intrinsic ambiguity of the pole-mass definition, and from the current experimental determinations. 
It may also be advantageous to employ other top-quark mass renormalization schemes, in order to assess the scheme dependence of the predictions. 
This strong sensitivity suggests that, in the 
long term, improved measurements of the $M_{t\bar{t}}$ spectrum, particularly with 
bin sizes of about $10$ to $20~\mathrm{GeV}$ at the HL-LHC, could also offer a promising new avenue for a more precise determination of $m_t$.


\subsection*{Acknowledgements}
We are grateful to various 
members of the 
ATLAS and 
CMS collaborations, in particular Christian Schwanenberger, Alexander Grohsjean, Laurids Jeppe, Afiq Anuar, for stimulating this work and for having shared with us various details of their experimental analyses. Part of the related discussions have taken place in the framework of the LHC Top Physics Working Group.\footnote{\url{https://indico.cern.ch/event/1608879/}.}
M.S. would like to thank for the hospitality at
the University of Hamburg while finalizing
this work.

The work of M.V.G. and S.-O.~M. has been supported in part by the Deutsche Forschungsgemeinschaft through the Research Unit FOR 2926, \emph{Next  Generation perturbative QCD for Hadron Structure: Preparing for the EIC}, project number 40824754.
The work of G.~L. has been supported by the Alexander von Humboldt foundation. 
The work of M.S. was supported by the Deutsche
Forschungsgemeinschaft under grant 396021762 -- TRR
257 \emph{Particle Physics Phenomenology after the Higgs Discovery}. 
The work of O. Z. has been supported by the
\emph{MSCA4Ukraine Programme} of the European Commission through the Alexander von Humboldt
foundation.


\appendix

\section{NRQCD cross sections with uncertainties}
\label{app:XSunc-xtd}

Here we report in Tabs.~\ref{tab:NLO_uncertainties} and~\ref{tab:NLONLL_uncertainties} the values of $\sigma^{\rm{NRQCD}}$ evaluated up to NLO and up to NLO plus NLL threshold resummation, respectively, for different  $M_{t\bar{t}}$ integration ranges, quoting explicitly the uncertainty obtained from the variation of each input parameter.
\begin{table}[h!]
\footnotesize
    \centering
    \begin{tabular}{|c|c|c|c|c|c|c|c|}
    \hline
    $M_{t\bar{t}}$ range &
    $\sigma^{\rm{NRQCD}}_{\rm{NLO}}$ &
    $\Delta_{\sigma}^{\mu_{R,0},\,\mu_{F,0}}$ &
    $\Delta_\sigma^{\mu_s}$ &
    $\Delta_\sigma^{\alpha_s}$ &
    $\Delta_\sigma^{\rm{PDFs}}$ &
    $\Delta_\sigma^{\Gamma_t}$ &
    $\Delta_\sigma$\\
     (GeV) &
    (pb) &
    (pb)&
    (pb)&
    (pb)&
    (pb) &
    (pb) &
    (pb)\\
    \hline
    340 - 345  &      4.42    &  $^{+0.36}_{-0.40}$    &  $^{+0.41}_{-0.28}$     & $^{+0.26}_{-0.25}$& $^{+0.16}_{-0.23}$&$^{+0.03}_{-0.02}$&$^{+0.62}_{-0.60}$\\
    \hline
340 - 346  &      5.61    &  $^{+0.46}_{-0.52}$&$^{+0.47}_{-0.34}$&$^{+0.30}_{-0.29}$&$^{+0.19}_{-0.30}$&$^{+0.04}_{-0.03}$&$^{+0.75}_{-0.74}$\\
\hline
340 - 347  &      6.92    &$^{+0.58}_{-0.64}$&$^{+0.55}_{-0.40}$&$^{+0.34}_{-0.33}$&$^{+0.22}_{-0.36}$&  $^{+0.04}_{-0.03}$&$^{+0.90}_{-0.90}$\\
\hline
340 - 348  &      8.38    &$^{+0.72}_{-0.79}$&$^{+0.63}_{-0.45}$&$^{+0.39}_{-0.37}$&$^{+0.25}_{-0.44}$ &$^{+0.04}_{-0.03}$&$^{+1.06}_{-1.08}$\\
\hline
340 - 349  &      9.97    &$^{+0.88}_{-0.95}$&$^{+0.71}_{-0.50}$& $^{+0.44}_{-0.42}$&$^{+0.27}_{-0.52}$&$^{+0.05}_{-0.04}$&$^{+1.24}_{-1.27}$     \\
\hline
340 - 350  &     11.67    &  $^{+1.05}_{-1.12}$&$^{+0.79}_{-0.55}$&$^{+0.49}_{-0.47}$&$^{+0.30}_{-0.61}$&$^{+0.05}_{-0.04}$&$^{+1.43}_{-1.47}$\\
\hline
340 - 351  &     13.47    &$^{+1.24}_{-1.31}$&$^{+0.86}_{-0.60}$&$^{+0.54}_{-0.53}$&$^{+0.32}_{-0.70}$& $^{+0.05}_{-0.04}$&$^{+1.64}_{-1.69}$\\
\hline
340 - 352  &     15.37    & $^{+1.44}_{-1.51}$&$^{+0.94}_{-0.65}$&$^{+0.60}_{-0.59}$&$^{+0.33}_{-0.80}$&$^{+0.05}_{-0.04}$&$^{+1.85}_{-1.92}$\\
\hline
340 - 353  &     17.34 & $^{+1.65}_{-1.71}$&$^{+1.02}_{-0.69}$&$^{+0.66}_{-0.64}$&$^{+0.35}_{-0.90}$&$^{+0.05}_{-0.04}$&$^{+2.08}_{-2.15}$\\
\hline
340 - 354  &     19.39    & $^{+1.86}_{-1.93}$&$^{+1.09}_{-0.74}$&$^{+0.72}_{-0.71}$&$^{+0.36}_{-1.00}$&$^{+0.05}_{-0.04}$&$^{+2.31}_{-2.40}$\\
\hline
340 - 355  &     21.50    &$^{+2.09}_{-2.15}$&$^{+1.17}_{-0.78}$&$^{+0.79}_{-0.77}$&$^{+0.38}_{-1.11}$&$^{+0.05}_{-0.04}$&$^{+2.55}_{-2.66}$\\
\hline
    \end{tabular}
    \caption{NLO predictions for $\sigma^{\rm{NRQCD}}$ with $m_t=172.5$ GeV, within a specific $M_{t\bar{t}}$ integration range, with the uncertainty $\Delta_\sigma^X$, determined from the change in $\sigma^{\rm{NRQCD}}$ when the input parameter $X$ is varied. In the last column the uncertainty $\Delta_\sigma$, obtained from the sum in quadrature of all the $\Delta_\sigma^X$ values is reported.} 
    \label{tab:NLO_uncertainties}
\end{table}

\begin{table}[h!]
    \footnotesize
     \centering
    \begin{tabular}{|c|c|c|c|c|c|c|c|}
    \hline
     $M_{t\bar{t}}$ range &
    $\sigma^{\rm{NRQCD}}_{\rm{NLO+NLL}}$ &
    $\Delta_{\sigma}^{\mu_{R,0},\,\mu_{F,0}}$ &
    $\Delta_\sigma^{\mu_s}$ &
    $\Delta_\sigma^{\alpha_s}$ &
    $\Delta_\sigma^{\rm{PDFs}}$ &
    $\Delta_\sigma^{\Gamma_t}$ &
    $\Delta_\sigma$\\
     (GeV) &
    (pb) &
    (pb)&
    (pb)&
    (pb)&
    (pb) &
    (pb) &
    (pb)\\
    \hline
    340 - 345  &      4.57  &$^{ + 0.15}_{- 0.02}$ &$^{+ 0.41}_{ - 0.28}$ &$^{ + 0.26}_{- 0.25}$ &$^{+ 0.16}_{ - 0.23}$   &$^{+ 0.03}_{ - 0.02}$&$^{+0.53}_{-0.44}$       \\
    \hline
340 - 346  &      5.80  &$^{ + 0.20}_{- 0.03}$ &$^{+ 0.47}_{ - 0.34}$ &$^{ + 0.30}_{- 0.29}$ &$^{+ 0.19}_{ - 0.30}$   &$^{+ 0.04}_{ - 0.03}$&$^{+0.62}_{-0.53}$      \\
\hline
340 - 347  &      7.16  &$^{ + 0.26}_{- 0.05}$ &$^{+ 0.55}_{ - 0.40}$ &$^{ + 0.34}_{- 0.33}$ &$^{+ 0.22}_{ - 0.36}$   &$^{+ 0.04}_{ - 0.03}$&$^{+0.73}_{-0.63}$     \\
\hline
340 - 348  &      8.67  &$^{ + 0.34}_{- 0.07}$ &$^{+ 0.63}_{ - 0.45}$ &$^{ + 0.39}_{- 0.37}$ &$^{+ 0.25}_{ - 0.44}$   &$^{+ 0.04}_{ - 0.03}$&$^{+0.85}_{-0.74}$      \\
\hline
340 - 349  &     10.31  &$^{ + 0.42}_{- 0.11}$ &$^{+ 0.71}_{ - 0.50}$ &$^{ + 0.44}_{- 0.42}$ &$^{+ 0.27}_{ - 0.52}$   &$^{+ 0.05}_{ - 0.04}$&$^{+0.97}_{-0.85}$      \\
\hline
340 - 350  &     12.08  &$^{ + 0.51}_{- 0.14}$ &$^{+ 0.79}_{ - 0.55}$ &$^{ + 0.49}_{- 0.47}$ &$^{+ 0.30}_{ - 0.61}$   &$^{+ 0.05}_{ - 0.04}$&$^{+1.10}_{-0.96}$     \\
\hline
340 - 351  &     13.95  &$^{ + 0.61}_{- 0.19}$ &$^{+ 0.86}_{ - 0.60}$ &$^{ + 0.54}_{- 0.53}$ &$^{+ 0.32}_{ - 0.70}$   &$^{+ 0.05}_{ - 0.04}$&$^{+1.23}_{-1.08}$     \\
\hline
340 - 352  &     15.92  &$^{ + 0.72}_{- 0.23}$ &$^{+ 0.94}_{ - 0.65}$ &$^{ + 0.60}_{- 0.59}$ &$^{+ 0.33}_{ - 0.80}$   &$^{+ 0.05}_{ - 0.04}$&$^{+1.37}_{-1.21}$      \\
\hline
340 - 353  &     17.97  &$^{ + 0.84}_{- 0.28}$ &$^{+ 1.02}_{ - 0.69}$ &$^{ + 0.66}_{- 0.64}$ &$^{+ 0.35}_{ - 0.90}$   &$^{+ 0.05}_{ - 0.04}$&$^{+1.52}_{-1.34}$     \\
\hline
340 - 354  &     20.09  &$^{ + 0.96}_{- 0.34}$ &$^{+ 1.09}_{ - 0.74}$ &$^{ + 0.72}_{- 0.71}$ &$^{+ 0.36}_{ - 1.00}$   &$^{+ 0.05}_{ - 0.04}$&$^{+1.66}_{-1.47}$     \\
\hline
340 - 355  &     22.28  &$^{ + 1.08}_{- 0.40}$ &$^{+ 1.17}_{ - 0.78}$ &$^{ + 0.79}_{- 0.77}$ &$^{+ 0.38}_{ - 1.11}$   &$^{+ 0.05}_{ - 0.04}$&$^{+1.82}_{-1.61}$      \\
\hline
    \end{tabular}
    \caption{Same as Table~\ref{tab:NLO_uncertainties}, also accounting for NLL soft-gluons threshold resummation in $\sigma^{\rm{NRQCD}}$.}
    \label{tab:NLONLL_uncertainties}
\end{table}


\section{Cross sections for different top-quark mass values}
\label{app:further_tables}

In this section we present the results for NLO NRQCD $d\sigma/dM_{t\bar{t}}$ and the associated excess with respect to perturbative QCD predictions, for values of the top-quark mass, differing from the central value $m_t=172.5$~GeV for $\pm1$~GeV, listed in Tabs.~\ref{tab:mt1715excessmu1} and ~\ref{tab:mt1735excessmu1}.
\begin{table}[h!]
    \centering
{\footnotesize
\begin{tabular}{|c|c|c|c|c|}  
    \hline
    $M_{t\bar{t}}$ range 
        & $\sigma_{\rm{NLO}}^{\rm{NRQCD}}$ 
        & $\Delta_\sigma$
        & $\sigma^{\rm{excess}}$ w.r.t.  
        & $\sigma^{\rm{excess}}$ w.r.t. 
        \\
        (GeV) &  (pb)
&  (pb)
&   NLO $t\bar{t}$ stable (pb)
& rescaled \texttt{hvq}+$t\bar{t}$ decay (pb)
    \\
        \hline

340 - 345  &  5.94  &$^{+0.50}_{ - 0.55}$ &     4.24  &     2.31    \\ \hline
340 - 346  &  7.44  &$^{+0.64}_{ - 0.70}$ &     4.45  &     2.33    \\ \hline
340 - 347  &  9.08  &$^{+0.81}_{ - 0.87}$ &     4.60  &     2.28    \\ \hline
340 - 348  & 10.84  &$^{+0.99}_{ - 1.05}$ &     4.70  &     2.17    \\ \hline
340 - 349  & 12.71  &$^{+1.18}_{ - 1.24}$ &     4.76  &     2.01    \\ \hline
340 - 350  & 14.67  &$^{+1.39}_{ - 1.44}$ &     4.78  &     1.79    \\ \hline
340 - 351  & 16.71  &$^{+1.60}_{ - 1.66}$ &     4.78  &     1.53    \\ \hline
340 - 352  & 18.82  &$^{+1.83}_{ - 1.88}$ &     4.74  &     1.23    \\ \hline
340 - 353  & 21.01  &$^{+2.06}_{ - 2.11}$ &     4.68  &     0.89    \\ \hline
340 - 354  & 23.25  &$^{+2.30}_{ - 2.35}$ &     4.59  &     0.53    \\ \hline
340 - 355  & 25.55  &$^{+2.55}_{ - 2.59}$ &     4.47  &     0.12    \\ \hline

    \end{tabular}
}
    \caption{\label{tab:mt1715excessmu1} Values of the NLO NRQCD cross section and its uncertainty and excess w.r.t. to fixed-order prediction (NLO with top stable), and {\texttt{POHWEGBOX-hvq}} followed by top-quark decay
rescaled to NNLO~+~NNLL QCD in different $M_{t\bar{t}}$ intervals. All cross sections are evaluated using as input $m_t$~=~171.5~GeV. The central hard scale is fixed to $\mu_R=\mu_F = m_t = 171.5$~GeV.
}    
\end{table}
\begin{table}[h!]
    \centering
{\footnotesize
\begin{tabular}{|c|c|c|c|c|}  
    \hline
    $M_{t\bar{t}}$ range 
        & $\sigma_{\rm{NLO}}^{\rm{NRQCD}}$ 
        & $\Delta_\sigma$
        & $\sigma^{\rm{excess}}$ w.r.t.  
        & $\sigma^{\rm{excess}}$ w.r.t. 
        \\
        (GeV) &  (pb)
&  (pb)
&   NLO $t\bar{t}$ stable (pb)
& rescaled \texttt{hvq}+$t\bar{t}$ decay (pb)
    \\
        \hline

340 - 345  &      2.80  &$^{+0.24}_{ -0.26}$ &     2.80  &     1.35 \\ \hline
340 - 346  &      3.88  &$^{+0.32}_{ -0.36}$ &     3.88  &     1.96 \\ \hline
340 - 347  &      4.94  &$^{+0.41}_{ -0.46}$ &     4.94  &     2.39 \\ \hline
340 - 348  &      6.08  &$^{+0.51}_{ -0.56}$ &     5.47  &     2.62 \\ \hline
340 - 349  &      7.35  &$^{+0.62}_{ -0.69}$ &     5.77  &     2.68 \\ \hline
340 - 350  &      8.76  &$^{+0.76}_{ -0.83}$ &     5.97  &     2.63 \\ \hline
340 - 351  &     10.30  &$^{+0.91}_{ -0.98}$ &     6.11  &     2.49 \\ \hline
340 - 352  &     11.95  &$^{+1.08}_{ -1.15}$ &     6.20  &     2.29 \\ \hline
340 - 353  &     13.69  &$^{+1.26}_{ -1.33}$ &     6.26  &     2.04 \\ \hline
340 - 354  &     15.52  &$^{+1.45}_{ -1.52}$ &     6.29  &     1.73 \\ \hline
340 - 355  &     17.44  &$^{+1.65}_{ -1.72}$ &     6.28  &     1.38 \\ \hline

    \end{tabular}
}
    \caption{\label{tab:mt1735excessmu1} Same as Tab.~\ref{tab:mt1715excessmu1} but for cross sections valuated using as input $m_t$~=~173.5~GeV. The central hard scale is fixed to   $\mu_R=\mu_F = m_t = 173.5$~GeV.
}    
\end{table}


\newpage

\bibliographystyle{JHEP}
\bibliography{coulomb_par} 
\end{document}